\documentclass[a4paper,11pt]{article}
\usepackage{jheppub} 
\usepackage{comment}
\usepackage{natbib}
\usepackage{float}
\usepackage{siunitx}
\usepackage{graphicx}
\usepackage{gensymb}
\usepackage{amsmath, cases}
\usepackage[usenames]{color}
\usepackage{amssymb}
\usepackage{hyperref} 
\newcommand{\be}{\begin{equation}}
\newcommand{\ee}{\end{equation}}
\newcommand{\bea}{\begin{aligned}}
\newcommand{\eea}{\end{aligned}}
\newcommand{\pr}{\partial}

\newcommand{\bse}{\begin{subequations}}
\newcommand{\ese}{\end{subequations}}

\newcommand{\bmm}{\begin{multline}}
\newcommand{\emm}{\end{multline}}

\newcommand{\mi}{\mathrm{i}}

\title{Axion–Photon Conversion in FLRW with Primordial Magnetic Fields: Explaining the Radio Excess} 

\author[1]{Setabuddin,}
\emailAdd{setabuddin.ahmad@gmail.com}
\affiliation[1]{Physics and Applied Mathematics Unit, Indian Statistical Institute, 203 B.T. Road, Kolkata 700108, India}
\author[1]{Md Riajul Haque,}
\emailAdd{riaj.0009@gmail.com}
\author[2]{Rajesh Karmakar,}
\emailAdd{rajesh@shu.edu.cn}
\affiliation[2]{Department of Physics, Shanghai University, 99 Shangda Road, Shanghai, 200444, China}
\author[1]{Supratik Pal}
\emailAdd{supratik@isical.ac.in}

\abstract{We explore the possibility of axion–photon conversion as a common origin of two low-frequency anomalies: the isotropic radio excess (ARCADE2) and the deep global 21-cm absorption trough (EDGES). From the axion–photon action in an FLRW background with primordial magnetic fields (PMFs), we derive the scale-dependent conversion probability including plasma effects. Resonant conversion, arising when the axion mass matches the plasma-induced photon mass, produces soft photons in the MHz–GHz range. By modeling stochastic PMFs with amplitude $B_0$ and spectral index $n_{\rm B}$, we show that axion-like particles with mass $\sim 10^{-14}$–$10^{-12}\,\mathrm{eV}$ and nanogauss-level nearly scale-invariant PMFs  can explain both ARCADE2 and EDGES. Heating from PMF dissipation via ambipolar diffusion and turbulent decay reduces the 21-cm trough, shifting the viable parameter space. Our results stem from a consistent theoretical framework developed from first principles and a combined analysis of the radio excess and global 21-cm signal, while remaining consistent with CMB bounds on PMFs and $\Delta N_{\rm eff}$. We conclude that global 21-cm observations may offer potential sensitivity to axions, primordial magnetism, and dark-sector physics.}

\renewcommand{\thesection}{\arabic{section}}

\begin{document}
\maketitle
\flushbottom
\section{Introduction}
Over the past two decades, the cosmic microwave background (CMB) observations have emerged as one of our most precious tools to understand the early universe \cite{Komatsu:2022nvu, Planck:2018nkj, Hu:2001bc}. While the CMB spectrum above $60~\rm GHz$ aligns well with theoretical predictions \cite{Rosenberg:2022sdy, Krachmalnicoff:2015xkg}, several low-frequency measurements, most notably from the Absolute Radiometer for Cosmology, Astrophysics, and Diffuse Emission 2 (ARCADE2) balloon mission \cite{Fixsen2011} and the Experiment to Detect the Global Epoch of Reionization Signature (EDGES) \cite{Bowman:2018yin}, reveal anomalies that warrant further investigation. The ARCADE2 collaboration, supported by additional low-frequency radio observations such as those from the Long Wavelength Array (LWA)~\cite{Taylor:2012fx} and the Haslam map~\cite{Haslam:1982zz}, reported an excess in the isotropic sky brightness temperature over the standard CMB blackbody spectrum at frequencies between 22 MHz and 10 GHz \cite{Kogut:2009xv, Rogers:2008vh, haslam1981, reich1986, Roger:1999jy, Guzman:2010da, guzman2011all}. This excess is usually modeled by a power law fitting function, $T_b(\nu) = T_{\rm CMB} + T_1 \left(\frac{\nu}{1\,{\rm GHz}}\right)^\beta$, with spectral index $\beta \approx -2.6$ and normalization $T_1 \approx 1.06\,{\rm K}$ at $1~{\rm GHz}$ \cite{Fixsen2011, Seiffert:2009xs}. The amplitude of this signal far exceeds expectations from known extragalactic source populations \cite{Dowell:2018mdb, Bowman:2018yin}, and its spatial smoothness strengthens the case for a cosmological, rather than galactic, origin. \\
Adding to the puzzle, the EDGES experiment \cite{Bowman:2018yin} reported in 2018 a strong absorption trough in the global 21-cm signal, centered at $z \simeq 17$ ($\nu \simeq 78$ MHz), with a depth of $-0.5^{+0.2}_{-0.5}\,\mathrm{K}$, nearly twice as large as predicted by $\Lambda$CDM given standard astrophysical assumptions about early star formation and the thermal history of the intergalactic medium. This unexpectedly deep absorption conflicts with the minimum value allowed in the Standard Model of cosmology at 99\% C.L., suggesting either an enhanced background radiation field or cooler-than-expected hydrogen gas, both pointing toward new physics beyond the Standard Model of cosmology. The claim, however, sparked intense debate: re-analyses of the EDGES data questioned the robustness of the result, highlighting the need for unphysical foreground parameters and possible unmodeled systematics \cite{Hills:2018vyr}. A follow-up observation with SARAS-3 in 2021 did not confirm the feature and rejected the best-fit EDGES profile at 95.3\% confidence \cite{Singh:2021mxo}. More recently, the EDGES collaboration introduced a Bayesian framework to address foreground modeling biases, showing that the intrinsic model used in \cite{Hills:2018vyr} is biased, and reaffirmed consistency with their original 2018 result \cite{Sims:2022nwg,Sims:2025kre}. Meanwhile, the ongoing third phase of EDGES, operational since 2022, has released updated data that remain consistent with the initial detection. Despite these controversies, interest in the global 21-cm signal remains high since other low-frequency experiments, including the Large Aperture Experiment to Detect the Dark Ages (LEDA), the Probing Radio Intensity at high-Z from Marion (PRIZM) experiment, and more recently the Radio Experiment for the Analysis of Cosmic Hydrogen (REACH) continue to explore similar signals and provide complementary insights into the cosmic dawn epoch \cite{Price2018LEDA, Chiang:2020pbx, deLeraAcedo:2022kiu}.
\\
Various theoretical proposals have been put forward as explanations of these anomalous observational signals. For the $21$-cm  absorption trough, they broadly fall into two classes. The first involves cooling of baryonic gas through interactions with cold dark matter \cite{Barkana:2018lgd, Fialkov:2018xre, Munoz:2018pzp}, which, however, the standard $\Lambda$CDM cosmology struggles to accommodate \cite{Barkana:2018qrx, Berlin:2018sjs}, pointing to possible new physics at late times. The second focuses on modifying the radiation background via additional photon injection through mechanisms such as dark photon decays \cite{Pospelov:2018kdh, Fraser:2018acy}, cosmic strings \cite{Brandenberger:2019lfm}, light dark matter, or axion-like particles (ALPs) \cite{Moroi:2018vci, Choi:2019jwx}. While photon injection alone does not deepen the $21$-cm absorption (and in fact makes it shallower), it can enhance the observed contrast by raising the effective radiation background, thereby potentially linking the EDGES trough to the radio excess seen by ARCADE2.  Coming to ARCADE2 and other low-frequency measurements \cite{Kogut:2009xv, Rogers:2008vh, haslam1981, reich1986, Roger:1999jy, Guzman:2010da, guzman2011all}, the frequency range associated with the observed excess in brightness temperature also encompasses the $78~\text{MHz}$ absorption trough reported by EDGES. This gives rise to a natural question of whether a unified source of soft photon injection could account for both signals. While some studies suggest mechanisms such as Bremsstrahlung heating \cite{Acharya:2023ygd} or relic neutrino decay \cite{Chianese:2018luo, Dev:2023wel}, these approaches typically fail to reproduce the full EDGES signal. A particularly intriguing scenario involves dark matter decaying into dark photons in the presence of a thermal dark photon bath \cite{Fornengo:2011cn}. These non-thermal dark photons can convert into ordinary photons near the epoch of reionization, producing a radio excess with a characteristic spectrum $T(\nu) \propto \nu^{-2.5}$ \cite{Berlin:2018sjs, Acharya:2022vck}. This conversion process could simultaneously enhance the apparent $21$-cm absorption depth. However, detailed analyses such as Ref.~\cite{Acharya:2022vck} show that these models are under significant tension with cosmological constraints. In particular, the dark photon explanation is challenged by CMB spectral distortion bounds, and future missions such as PIXIE could critically test its predictions. Moreover, the model struggles to explain the spatial smoothness of the radio background, as best-fit parameters imply significant anisotropies from clustered dark matter at low redshift.\\
In light of these challenges, alternative mechanisms that can generate late-time soft photon backgrounds without thermalizing in the presence of plasma or violating spectral constraints become highly attractive. Photon injections through axion oscillations \cite{Moroi:2018vci, Choi:2019jwx} provide such a mechanism.  
\textit{Axions or Axion-like particles (ALPs)} arise in many extensions of the Standard Model of particle physics. These ALPs being pseudo-scalar particles, couple with photon as $\mathcal{L} \supset g_{\phi\gamma} \phi F_{\mu\nu} \tilde{F}^{\mu\nu}$, and get converted into photons in the presence of large-scale magnetic fields \cite{Raffelt:1987im, Mirizzi:2007hr, Caputo:2019tms, Addazi:2024mii}. Moreover, if such phenomena happen in the presence of a plasma medium, photons acquire an effective mass $m_\gamma(z)$, which decreases with redshift. When this matches the ALP mass $m_\phi$, resonant conversion can occur, leading to the production of soft photons without requiring any particle decay or energy injection into the plasma. This resonant axion-photon conversion has several appealing features. It can take place after recombination, avoiding strong early-universe bounds; it does not inject entropy or ionize the gas; and it produces a low-frequency photon population that can accumulate over time.  Such a mechanism has been previously implemented \cite{Moroi:2018vci} to investigate the $21$-cm trough as measured in EDGES. Recently, it has been found out \cite{Addazi:2024mii} that the same mechanism can accommodate the excess in the radiation background as observed by ARCADE2 and other low-frequency observations alongside EDGES. Particularly the authors have shown that ALPs with masses $m_\phi \sim 10^{-14} - 10^{-12} \, \text{eV}$, coupling strengths $g_{\phi\gamma} \sim 10^{-13} - 10^{-11} \, \text{GeV}^{-1}$, and coherent magnetic fields of nanogauss strength can lead to resonant conversion of axion into photons at certain redshift, that generates the excess in the background spectrum. 
\\
In this work, we build upon the idea of axion-photon conversion \textit{starting from first principles in an expanding universe}. We derive the equations of motion from the axion--photon action in an FLRW background, where the scale factor naturally enters the dynamics but has only been partially considered in previous studies~\cite{Raffelt:1987im, Mirizzi:2007hr, Higaki:2013qka, Caputo:2019tms, Addazi:2024mii}. Properly incorporating the expanding background, we compute the resonant conversion probability in the presence of a stochastic primordial magnetic field and show that, even with spacetime expansion, axion--photon mixing can be analyzed perturbatively. We include realistic models of magnetic fields with scale-invariant and nearly scale-invariant power spectra, as well as evolving plasma effects that determine the photon effective mass, and demonstrate that axion--photon conversion can naturally produce a population of soft photons peaking in the MHz--GHz range, sufficient to explain the ARCADE2 excess and enhance the background temperature relevant for the EDGES absorption. The EDGES result indicates that the gas kinetic temperature in the intergalactic medium (IGM) was cooler than the CMB temperature at $15 \lesssim z \lesssim 20$~\cite{Bowman:2018yin}, and this temperature can be significantly influenced by primordial magnetic fields, which act as additional cosmological heating sources through ambipolar diffusion and decaying turbulence in the late universe. Several works have shown that measurements of the global 21-cm signal constrain the amplitude and scale dependence of PMFs~\cite{Schleicher:2008hc, Sethi:2009dd, Bhaumik:2024efz}. Following these, we evaluate the global thermal history across the PMF parameter space ($B_0, n_B$), finding that the scale-invariant case ($n_B = -3$) does not contribute significantly to magnetic heating, whereas the nearly scale-invariant case ($n_B = -2.75$) is strongly affected by PMF dissipative heating during the post-recombination era. To bring these observations on the same footing, we investigate the relevant parameter space and find that our results are significant different from the previous works done in this direction~\cite{Addazi:2024mii}. We attribute this to a consistent development of the theoretical framework starting from the first principles as well as a consistent, combined analysis with both the radio excess and 21-cm global signal observations, keeping consistency with CMB bounds on PMF and $\Delta N_{\rm eff}$. We map out the viable ALP parameter space consistent with existing bounds and show that this mechanism succeeds where dark photon models often fail, particularly in avoiding CMB spectral distortions~\cite{Arsenadze:2024ywr, Chluba:2024wui, aramburo2024dark} and $\Delta N_{\rm eff}$ constraints~\cite{ACT:2025tim, ACT:2020gnv, Rossi:2014nea, mangano2011robust}.
\\
This paper is organized as follows. In Section~\ref{sec-2}, we formulate the axion--photon system in an expanding FLRW background and derive the evolution of the mixing parameters. Section~\ref{sec-3} introduces the modeling of stochastic primordial magnetic fields and presents the redshift-dependent expression for the conversion probability. In Section~\ref{sec-4}, we connect resonant conversion to the radio brightness temperature and compare our results with ARCADE2 and other low-frequency radio observations. In Section~\ref{sec-5}, we apply our framework to the global 21-cm absorption signal reported by EDGES, identifying benchmark parameter choices consistent with the anomaly and discussing the impact of PMF-induced heating on the allowed parameter space. Finally, Section~\ref{sec-6} concludes with a summary of our main findings and outlines the prospects for testing this framework with upcoming 21-cm experiments, radio surveys, and next-generation CMB missions.

%%%%%%%%%%%%%%%%%%%%%%%%%%%%%%%%%%%%%%%%%%%%%%%%
\section{Axion-photon conversion in the FLRW spacetime with background magnetic field}\label{sec-2}
The large-scale isotropy and homogeneity of the universe uniquely determine the spacetime geometry, which we describe using the conformally flat Friedmann–Lemaître-Robertson-Walker (FLRW) metric in conformal time \cite{Baumann:2022mni}.\footnote{The metric takes the form 
$ds^{2}=a^{2}(\eta)\left(-d\eta^{2}+d\mathbf{x}^{2}\right)$, 
where $a(\eta)$ is the scale factor and $\mathbf{x}$ are the comoving spatial coordinates.}
During its early evolution, small quantum fluctuations in the universe may have introduced curvature perturbations. Particularly, in the inflationary paradigm \cite{guth1981inflationary, linde1982new}, these small-scale modes grow as they re-enter the horizon and are understood to be forming the present structure of our universe\cite{mukhanov1981quantum}. In addition to the standard model particles produced in the early universe, a variety of ultralight particles may have also emerged. Among these, axions naturally arise from the Peccei–Quinn mechanism \cite{peccei1977cp} proposed to solve the strong CP problem and carry important cosmological implications. There are various production mechanisms of axions in the early universe, such as through misalignment \cite{preskill1983cosmology}, radiation from cosmic strings \cite{yamaguchi1999evolution} and domain walls \cite{sikivie1982axions}, thermal scatterings \cite{graf2011thermal}, etc. Being a pseudo-scalar particle, the axion couples with the electromagnetic field in a certain way. The action for the axion-photon interaction, minimally coupled with the background spacetime, can be expressed as \cite{Carosi:2013rla},
\be
\mathcal{A}=\int \sqrt{-g}d^4x\left[-\frac{1}{2}(\pr_\mu \phi\pr^\mu \phi+m^2_\phi \phi^2)-\frac{1}{4}F_{\mu\nu}F^{\mu\nu}-J_\mu A^\mu-\frac{1}{4}g_{\phi\gamma}\phi F_{\mu\nu}\tilde{F}^{\mu\nu}\right],
\ee
where $g_{\phi\gamma}$ represents the axion-photon coupling constant. Whereas, $\sqrt{-g}$ denotes the determinant of the spacetime metric, $\tilde{F}^{\mu\nu}=\epsilon^{\mu\nu\rho\sigma}F_{\rho\sigma}/2$ represents the dual of the electromagnetic field-strength tensor, $F_{\mu\nu}$, and $\phi$, a pseudo scalar field, representing the axion. To incorporate the interaction of the photon with the plasma medium, we have added the source term with $J^\mu$ denoting the four-current density \cite{landau1975classical}. Notably, considering plasma medium with the covariantly conserved current, $\nabla_\mu J^\mu=0$, makes sure that the action is gauge invariant. By varying the action we obtain the following set of coupled equations, 
\be\label{em.sc.eq}
\bea
\frac{1}{\sqrt{-g}}\pr_\mu({\sqrt{-g}}g^{\mu\alpha}g^{\nu\beta}F_{\alpha\beta})+J^\nu-g_{\phi\gamma}\tilde{F}^{\mu\nu}\pr_\mu \phi &=0,\\
\frac{1}{\sqrt{-g}}\pr_\mu({\sqrt{-g}}g^{\mu\nu}\pr_\nu\phi)-m^2_\phi \phi-\frac{1}{4}g_{\phi\gamma}F_{\mu\nu}\tilde{F}^{\mu\nu} &=0.
\eea
\ee
It is interesting to note that the coupling term appears as the source in the equation of motion and is mainly responsible for axion--photon conversion. In the subsequent analysis, we follow the convention of \cite{Turner:1987bw}, where the electromagnetic field tensor $F_{\mu\nu}$ is expressed in terms of the physical electric and magnetic fields in the conformally flat FLRW background \footnote{Explicitly, the components of the electromagnetic field tensor take the form
\[
F_{0i} = -a^2(\eta)\,E_i, 
\qquad 
F_{ij} = a^2(\eta)\,\epsilon_{ijk} B^k,
\]
where \(E_i\) and \(B_i\) denote the physical electric and magnetic field components.} Substituting this decomposition, the governing equations can be expressed in terms of the electric and magnetic fields as  
\begin{align}\label{field.eqn.turner}
-\frac{1}{a^2}\,\pr_\eta\!\left(a^2 \mathbf{E}\right) + \nabla\times \mathbf{B} + a^2 \mathbf{J}
   - g_{\phi\gamma}\!\left[\dot{\phi}\,\mathbf{B} + \nabla \phi \times \mathbf{E}\right] 
   &= 0, \\
\pr^2_\eta \phi + 2\frac{\pr_\eta a}{a}\pr_\eta \phi - \nabla^2 \phi 
   + g_{\phi\gamma}\,a^2\,\mathbf{E}\cdot\mathbf{B} 
   &= 0.
\end{align}
Here $\nabla$ represents the 3-dimensional gradient operator in Cartesian coordinates. Whereas, ${\bf J}=\sigma {\bf E}$, denotes the spatial current density with $\sigma$ representing the conductivity of the medium. To move forward, we make the following gauge choice, ${\bf \nabla}\cdot{\bf A}=0, A_0=0$, known as the radiation gauge. Fixing the gauge, the above equations read as,
\be
\bea
\pr^2_\eta {\bf A}+a^2(\eta){\bf \nabla}\times {\bf B}-a^2(\eta)\sigma\pr_\eta{\bf A}-g_{\phi\gamma}a^2(\eta){\bf B} \pr_\eta\phi+g_{\phi\gamma}{\bf \nabla} \phi\times \pr_\eta{\bf A} &=0,  \\
\pr^2_\eta \phi+\frac{2\pr_\eta a(\eta)}{a(\eta)}\pr_\eta \phi-{\bf \nabla}^2 \phi+m^2_\phi a^2(\eta)\phi-g_{\phi\gamma}{\bf B}\cdot \pr_\eta{\bf A} &=0,
\eea
\ee
where we have used ${\bf E}=-\pr_\eta{\bf A}/a^2$.
%where, we followed the definition, $E_\mu=F_{\mu\nu}u^\nu$ and $B_\mu=\epsilon_{\mu\alpha\beta}F^{\alpha\beta}/2$, with the four-velocity (time-like) of the observer given by $u^\mu$, satisfying $u^\mu u_\mu=-1$. 
We will now linearize these equations about some background magnetic field solutions (${\bf B}_0$), 
\be
{\bf B}\to {\bf B}_0+\frac{1}{a^2(\eta)}{\bf \nabla}\times{\bf A}(\eta, l),
\ee
where the scale factor appears due to the convention. We choose the $l$-axis to lie along the propagation direction, i.e., a spatial coordinate perpendicular to the $x$-$y$ plane. We reserve $z$ to denote the redshift parameter later in our discussion.  Substituting the above decomposition, the equations of motion, up to linear order in fluctuations of the fields, turn out as, 
\be
\bea
\pr^2_\eta{\bf A}(\eta,l) - \pr^2_l {\bf A}(\eta,l) - a^2(\eta)\sigma\pr_\eta{\bf A}- g_{\phi\gamma}a^2(\eta){\bf B}_0\pr_\eta {\phi}(\eta, l) &= 0, \\
\pr^2_\eta\phi(\eta, l) + 2 \mathcal{H}(\eta) \pr_\eta \phi(\eta, l) - \pr^2_l \phi(\eta, l) +m^2_\phi a^2(\eta)\phi(\eta, l) + g_{\phi\gamma}{\bf B}_0\cdot \pr_\eta{\bf A}(\eta, l) &= 0,
\eea
\ee
where $\mathcal{H}\equiv\pr_\eta a(\eta)/a(\eta)$ represents the conformal Hubble parameter. Given the propagation of the fluctuations along $l$-direction, we find it convenient to consider the following form of the fluctuations,
\be
\bea
{\bf A} (\eta, l) &\sim \mi{\begin{pmatrix}A^x(\eta, l) \\ A^y(\eta, l)
\end{pmatrix}}e^{-\mi k l}, \\
\phi(\eta, l) &\sim  \Tilde{\phi}(\eta, l) e^{-\mi k l},
\eea
\ee
with $k$ denoting the linear momentum along $l$ (in natural unit). On top of this generic decomposition, we make the following assumption: the background magnetic field is varying on a much larger scale (as our analysis will be based on a magnetic field model that is large-scale, isotropic, and homogeneous), than the axion-photon fluctuation \cite{Raffelt:1987im}, which implies, 
\be\label{freq.length}
\bea
k {A^x}(\eta, l) &> \mi \pr_l{A^x}(\eta, l),\\
k {A^y}(\eta, l) &> \mi \pr_l{A^y}(\eta, l),\\
k {\Tilde{\phi}}(\eta, l) & >\mi \pr_l \Tilde{\phi}(\eta, l).
\eea
\ee
In terms of the coherent length of the background magnetic field, this translates to $k>l^{-1}_{\rm coherent}$. The comoving coherent length is usually considered to be $l_{\rm coherent}\sim 1{\rm Mpc}$ \cite{Moroi:2018vci}. Importantly, we only consider the leading order term for the spatial variation, attributing more importance to the time variation given the local expansion of spacetime. Furthermore, considering high frequency fluctuation in time we linearize the time variation of the fluctuation up to first order, $ \pr_\eta^2 + k^2 =  (\mi \pr_\eta + k)(-\mi \pr_\eta +k)=(\mi \pr_\eta +k)(\omega+k) \simeq 2\omega (\mi \pr_\eta + \omega)$. Finally, these equations are expressed below in a matrix form:
\be\label{meq}
\mi\pr_\eta \begin{pmatrix}
A^x\\
A^y\\
\Tilde{\phi}
\end{pmatrix}=\begin{pmatrix}
        -\omega-\frac{\omega^2_{pl} a^2(\eta)}{2\omega} & 0 & -\frac{1}{2} g_{\phi \gamma} a^2(\eta) B^x_{0} (1 + \frac{\mi \mathcal{H}(\eta)}{\omega}) \\
        0 & -\omega-\frac{\omega^2_{pl} a^2(\eta)}{2\omega} & -\frac{1}{2} g_{\phi \gamma} a^2(\eta) B^y_{0}\left(1 + \frac{\mi \mathcal{H}(\eta)}{\omega}\right) \\
        \frac{\mi}{2} g_{\phi \gamma}  B^x_{0} (1 + \frac{\mi \mathcal{H}(\eta)}{\omega}) & \frac{\mi}{2} g_{\phi \gamma} B^y_{0} (1 + \frac{ \mi \mathcal{H}(\eta)}{\omega}) & -\omega-\frac{m^2_\phi a^2(\eta)}{2\omega}+\mi\mathcal{H}(\eta)
    \end{pmatrix}
    \begin{pmatrix}
A^x\\
A^y\\
\Tilde{\phi}
\end{pmatrix}.
\ee
In obtaining the above equation, we have kept upto $\mathcal{O}(1/\omega)$, having considered high frequency axion-photon fluctuation. We define the axion-photon conversion probability as
\be
\mathcal{P}(\eta,l)=\frac{\left|A^x(\eta,l)\right|^2+\left|A^y(\eta,l)\right|^2}{|\Tilde{\phi}(\eta_0,l)|^2},
\ee
where $\eta_0$ denotes the initial conformal time, which we consider at the moment of recombination.
From the matrix form of the equations of motion, it is obvious that the off-diagonal term would mix the components, thereby leading to the conversion of axion-photon \cite{Raffelt:1987im}. Due to mixing, solving the equation of motion analytically for computing the conversion probability is nontrivial. Nevertheless, the computation can be carried out perturbatively, with the coupling constant, $g_{\phi\gamma}$, serving as the perturbation parameter. Considering up to first order of the coupling constant in the expansion of the axion-photon fluctuation, the conversion probability can be expressed in integral form as (see Appendix.\ref{derive.prob} for the details of the derivation), 
\be\label{prob.eta}
\bea
\mathcal{P}(\eta,\Delta l) =&\frac{g_{\phi \gamma}^2}{4}
\int_{\eta_0}^{\eta} d\eta_1 \int_{\eta_0}^{\eta} d\eta_2 \int \frac{d^3{\rm k}}{\rm (2\pi)^3} e^{\mi{\rm k} \cos \theta \Delta l} P_B({\rm k}) \left(1 + \frac{\mi \mathcal{H}(\eta_1)}{\omega_1} - \frac{\mi \mathcal{H}(\eta_2)}{\omega_2} \right) \\
&~~~~~~~~~~~~~~~~~~~~~~~~~~~~~~~\times e^{-\mi \int_{\eta_1}^{\eta_2} d\eta_3 [\Delta_{pl}(\eta_3)-\Delta_\phi(\eta_3)]}\, ,\\
\eea 
\ee
where, $\Delta_\phi(\eta) =m_{\phi}^2 a^2(\eta)/(2\omega)$, and $\Delta l=(l_1 - l_2)$. Whereas, $\Delta_{pl}=\omega^2_{pl}a^2(\eta)/(2\omega)$, with the plasma frequency $\omega_{pl}$. Interestingly, the photon acquires an effective mass in the plasma medium and can be expressed as $m_\gamma\sim \omega_{pl}=\sqrt{e^2n_e/m_e}$. Here, $e$ is the electron charge, $m_e$ is the electron mass. The electron number density is given by $n_e=n_{b0}(1+z)^3x_e(z)$, where $n_{b0}$ denotes the present-day baryon number density, $z$ is the redshift, and $x_e(z)$ is the ionization fraction as a function of redshift. Note that the wave number, ${\rm k}$  in the magnetic power spectrum, $P_B({\rm k})$, is assigned a different sign for the background magnetic field compared to that used for axion–photon propagation. From the above expression, it is clear that the axion-photon probability depends not only on the propagated comoving path, $\Delta l$, but also on the comoving conformal time. This is a natural consequence in an expanding spacetime \cite{Moroi:2018vci, Addazi:2024osi}. Before proceeding to the actual evaluation, it is also important to realize that apart from the momentum integral, which is determined by the background magnetic field model, the essential behaviour of the conversion probability depends on the exponentially oscillating integrand. This oscillation, interestingly, gets extremized when the effective mass of the photon, $m_\gamma$, becomes equal to the axion mass, $m_\phi$. We will return to this point in the next section once the integral is written in terms of the redshift factor. 
%%%%%%%%%%%%%%%%%%%%%%%%%%%%%%%%%%%%%%%%%%%%%%%%%%%%%%%%%%%%%%%%%%%%%%%%%%%%%%%%%%%%%%%%%%%%%%%%%%%%%%%%
\section{Primordial magnetic field modeling and conversion probability}\label{sec-3} 
Observed large scale intergalactic magnetic fields could plausibly trace back to primordial origins. In the early universe, extremely weak primordial magnetic fields may have been generated during inflation through quantum fluctuations in form of seed fields \cite{ratra1992cosmological}. Importantly, these seed fields arise when conformal invariance of electromagnetism is broken, for example via explicit time dependent coupling of the form: $\mathcal{L}\subset I^2(\eta)F_{\mu\nu}F^{\mu\nu}$, with $I=(a/a_{\rm end})^n$ for $a\leq a_{\rm end}$ and $I=1$ for $a >a_{\rm end}$, $a_{\rm end}$ representing the scale factor at the end of the inflation \cite{Kobayashi_2019}. Although this prescription takes care of the coupling, motivated by the phenomenology during and after inflation through the scale factor, moreover, it could also be directly considered as a function of the inflaton field \cite{Vilchinskii:2017qul, Sharma:2018kgs, Sharma:2017eps,Tripathy:2021sfb, Ferreira:2013sqa,Haque:2020bip}. Similar scalar field-induced breaking has also been studied, motivated by a dilaton \cite{Bamba:2004cu, Gasperini:1995dh}, or spectator-like scalar field \cite{Fujita:2016qab}, etc. On the other hand, higher curvature coupling, such as $\sim RF_{\mu\nu}F^{\mu\nu}$ or with higher powers of $R$, the Ricci scalar-curvature, also potentially breaks the conformal invariance and helps to generate such a large-scale magnetic field \cite{Bamba:2020qdj}. To assess whether inflationary magnetogenesis can account for today’s large-scale magnetic fields, it is crucial to examine the spectral properties of the generated fields. In particular, a nearly scale-invariant magnetic power spectrum is required so that magnetic fields survive on cosmological scales \cite{Bonvin:2013tba}. The conditions for scale invariance can be derived directly from the mode equation of the gauge field, particularly depending upon the form of conformal-breaking coupling \cite{Kobayashi_2019}. 
\\
We begin with the consideration of a scale-invariant power spectrum for a stochastic magnetic field from post-inflationary magnetogenesis \cite{Durrer:2013pga}. The form of the scale invariant power spectrum of the primordial magnetic field is usually expressed as, 
\be
P_B({\rm k})=\frac{\pi^2 B^2_0}{{\rm k}^3},
\ee
where $B_0$ is the amplitude of the magnetic field, normalized at $1 {\rm Mpc}$ length scale \cite{Minoda:2018gxj}. The average strength of the magnetic field can be obtained from the correlation as, $\sqrt{\bf B^2(x)}$, which for such a scale invariant spectrum diverges in both the UV and IR limit. We regularize this by introducing the UV-IR cutoff scale, ${\rm k}_{IR}<{\rm k}<{\rm k}_{UV}$. We set the IR limit utilizing the coherent length scale, so that ${\rm k}_{IR}\sim 1 {\rm Mpc}^{-1}$ \cite{Brandenburg:2018ptt}. The UV cutoff scale, ${\rm k}_{UV}$, is discussed below.

 \subsection{Nearly scale invariant magnetic spectrum} Allowing for deviations from an exactly scale-invariant power spectrum expands the parameter space in which axion–photon conversion can remain compatible with the ARCADE-2 low-frequency excess and the EDGES observations. The power spectrum for the nearly scale invariant primordial magnetic field is often expressed in the following manner \cite{Kahniashvili_2010, Minoda:2018gxj},  
\be
    P_B({\rm k})=\frac{(2\pi)^{n_B+5}}{\Gamma\left(\frac{n_B+3}{2}\right)}B_0^2\frac{{\rm k}^{n_B}}{{\rm k}^{n_B+3}_n}, 
\ee
where $n_B$ denotes the magnetic spectral index. Notably, $n_B=-3$ provides for the scale invariant ({\rm SI}) primordial magnetic power spectrum, whereas for a nearly scale invariant ({\rm NSI}) spectrum we consider $n_B=-2.75$ for the present analysis. Whereas $B_0$, in the above expression, represents the amplitude set in the case for the SI magnetic power spectrum, which we set to vary. Specifically, we choose $B_0\in [0.1, 1.0]{\rm nG}$. On the other hand we keep $k_n$, a reference scale assisting $k$, fixed at $2\pi~{\rm Mpc}^{-1}$.
\\
Nevertheless, in order to deal with this large scale magnetic field power spectrum we utilize the following approximate cutoff scale \cite{Kahniashvili_2010},
\be\label{kD}
{\rm k}_D\simeq 10^2\sqrt{\frac{(2\pi)^{n_B+3}h}{\Gamma(n_B/2+5/2)}}\left(\frac{1{\rm nG}}{B_0}\right){\rm Mpc}^{-1}\,,
\ee
where $h$ denotes the dimensionless Hubble constant, defined as $h\equiv H_0/(100~{\rm km~s^{-1} Mpc^{-1}})$. Note that, in \cite{Kahniashvili_2010}, effective magnetic field ($B_{\rm eff}$) has been used to define the cutoff, and is exactly same with $B_0$ for scale invariant spectrum. For nearly scale invariant case this quantity slightly deviates from $B_0$ and gets a wavenumber, ${\rm k}$, dependence. In our present analysis, given a small deviation from the scale invariant case, we treat them to be same. However, it is important to mention the BBN bound on this effective magnetic field $B_0\leq 8.4\times 10^2{\rm nG}$ \cite{Kahniashvili:2009qi}. Further tighter constraints on the upper bound can be obtained from latest CMB data from Planck 2018 as:
%Also from the measurement of CMB anisotropies, this constraint turns out to 
$B_0< 4.4 {\rm nG}$ \cite{Planck:2015zrl}. Notably, both of these constraints are at $1 {\rm Mpc}$ length scale. 
\\
With this setup for the background magnetic field, we first evaluate the momentum integral in the conversion probability \eqref{prob.eta},
\be
\int \frac{d^3{\rm k}}{(2\pi)^3} \ e^{\mi{\rm k} \cos \theta (l_1 - l_2)} P_B({\rm k})
=\left\{\begin{matrix}
\frac{1}{2\pi^2} \frac{(2\pi)^{n_B+5}}{\Gamma\left(\frac{n_B+3}{2}\right)}\frac{B_0^2}{{\rm k}^{n_B+3}_n}\int_0^{{\rm k}_{UV}} \frac{d{\rm k}}{{\rm k}^{-n_B-2}} \frac{\sin [{\rm k} (l_1 - l_2)]}{{\rm k} (l_1 - l_2)},~~~~~~~{\rm for~NSI}\\
\frac{1}{2}B^2_0\int_{{\rm k}_{IR}}^{{\rm k}_{UV}} \frac{d{\rm k}}{\rm k} \frac{\sin [{\rm k} (l_1 - l_2)]}{{\rm k} (l_1 - l_2)},~~~~~~~~~~{\rm for~SI}.
\end{matrix}
\right.
\ee
In the rest of the steps, we refer to this integral as follows,
\be
\bea
    \int \frac{d^3{\rm k}}{(2\pi)^3} \ e^{\mi{\rm k} \cos \theta (l_1 - l_2)} P_B({\rm k})=f({\rm k}_{IR}, {\rm k}_{UV}, B_0, \Delta l).
\eea
\ee
With this closed form expression, the conversion probability simplifies to 
\be
\bea
\mathcal{P}(\eta, \Delta l)&=\frac{g^2_{\phi \gamma}}{4}f({\rm k}_{IR}, {\rm k}_{UV}, B_0, \Delta l)
\int_{\eta_0}^{\eta} d\eta_1 \int_{\eta_0}^{\eta} d\eta_2 \left(1 + \frac{\mi \mathcal{H}(\eta_1)}{\omega_1} - \frac{\mi \mathcal{H}(\eta_2)}{\omega_2} \right)\times\\
&~~~~~~~~~~~~~~~~~~~~~~~~~~~~~~~~~~~~~~~~~~~~~~~~~~~~~~~~~~~~e^{-\mi \int_{\eta_1}^{\eta_2} d\eta_3 \left[\Delta_{pl}(\eta_3)-\Delta_\phi(\eta_3)\right]}.
\eea
\ee
In what follows, we will convert the above integral over conformal time $\eta$ to that over the redshift $z$, using the standard relations between $a$, $\eta$, and $z$. Explicitly, $1+z=1/a(\eta)$ and $d\eta=-dz/H(z)$.  Utilizing this transformation, we express the conversion probability as
\be
\bea
\mathcal{P}(z, \Delta l)&=\frac{g^2_{\phi \gamma}}{4}f({\rm k}_{IR}, {\rm k}_{UV}, B_0, \Delta l)\int_{z_0}^{z} \frac{dz_1}{H(z_1)} \int_{z_0}^{z}\frac{dz_2}{H(z_2)}\\
&~~~\times\left(1 + \frac{\mi H(z_1)}{\omega_0(1+z_1)^2} - \frac{\mi H(z_2)}{\omega_0(1+z_2)^2} \right)\exp\left[\mi \int_{z_1}^{z_2} \frac{dz_3}{H(z_3)} \{\Delta_{pl}(z_3)-\Delta_\phi(z_3)\}\right],
\eea
\ee
where the effect of expansion is taken care of. 
\begin{figure}[t]
\centering
\includegraphics[scale=0.6]{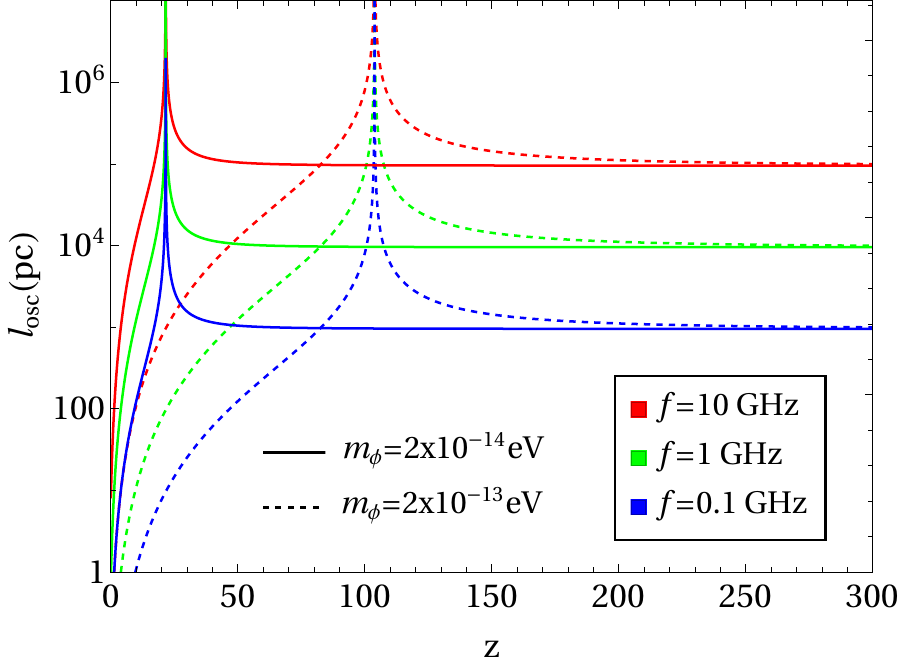}
\caption{The effective oscillation length scale has been plotted with the redshift factor, $z$, for different axion mass, $m_\phi$, at different frequency, $f=\omega_0/2\pi$.}\label{losc}
\end{figure}
Now we need to bring back our previously stated point regarding the behavior of the oscillating integral, specifically the exponential term in above equation. We see that on certain redshift, when  $\Delta_{pl}=\Delta_\phi$, the extremization happens. We can define an equivalent length scale associated to this oscillation \cite{Mirizzi:2007hr, Moroi:2018vci, Addazi:2024mii}, $l_{\rm osc}(z)\equiv|\Delta_{pl}(z)-\Delta_\phi(z)|^{-1}$, with the realization that $d\eta = - dz/H(z)$ as mentioned previously. In Fig.\ref{losc}, we have plotted this length scale in terms of the redshift factor. We notice that there appears to be a resonant increment of this scale at a certain redshift, depending on the axion mass, $m_\phi$.   
\begin{comment}
&=\frac{g^2_{\phi \gamma} P_{B0}}{2}f(k_{IR}, k_{UV}, \Delta l)\int_{z_0}^{z} \frac{1}{H(z_1)}dz_1 \int_{z_0}^{z}\frac{1}{H(z_2)} dz_2 \left(1 + \frac{i \mathcal{H}(\eta_1)}{\omega_1} - \frac{i \mathcal{H}(\eta_2)}{\omega_2} \right)\\
&~~~~~~~\exp\left[i\int_{z_1}^{z_2} \frac{1}{H(z_3)}dz_3 \{\Delta_{pl}(z_3)-\Delta_\phi(z_3)\}\right]\\   
\end{comment}
As per our consideration of high frequency oscillation for axion-photon fluctuation, $H/\omega<<1$, we express the above equation as
\be
\bea
\mathcal{P}(z, \Delta l)&=\frac{g^2_{\phi \gamma}}{4}f({\rm k}_{IR}, {\rm k}_{UV}, B_0, \Delta l)\int_{z_0}^{z} \frac{dz_1}{H(z_1)} \int_{z_0}^{z}\frac{dz_2}{H(z_2)} \\
&~~~~~~\times\exp\left[\frac{\mi H(z_1)}{\omega_0(1+z_1)^2} - \frac{\mi H(z_2)}{\omega_0(1+z_2)^2} + \mi \int_{z_1}^{z_2} \frac{dz_3}{H(z_3)} \{\Delta_{pl}(z_3)-\Delta_\phi(z_3)\}\right].
\eea
\ee
Before we proceed, we first evaluate the integral inside the argument of the exponential as,
\be
\int^{z_2}_{z_1}\frac{1}{H(z_3)}dz_3 \{\Delta_{pl}(z_3)-\Delta_\phi(z_3)\}
%&=\frac{1}{H_0\sqrt{\Omega_m}}\int^{z_2}_{z_1}\frac{1}{(1+z_3)^{3/2}}\left[\alpha-\frac{\beta}{(1+z_3)^3}\right]\\
=\frac{1}{H_0\sqrt{\Omega_m}}\left[-\frac{e^2n_{b0}x_e}{\omega_0m_e(1+z_3)^{\frac{1}{2}}}+\frac{m^2_\phi}{7\omega_0(1+z_3)^{7/2}}\right]^{z_2}_{z_1},
\ee
where we have used $H(z)=H_0\sqrt{\Omega_m}(1+z)^{3/2}$, with $\Omega_m$ denoting the matter energy density of the universe at present. Whereas, we have utilized the following expression for the terms involving the axion mass and plasma medium,   
\be
\bea
\Delta_{pl}(z) &= \frac{1}{2\omega_0}\frac{e^2n_{b0}x_e(z)}{m_e},\\
\Delta_\phi(z) &= \frac{m^2_\phi}{2\omega_0(1+z)^3}.
\eea
\ee
For simplicity, it has been assumed that the ionization fraction, $x_e(z)$ doesn't change significantly with redshift, even though in principle it might have a dynamical evolution with redshift. In the post-recombination era, the universe is predominantly neutral with a residual electron fraction of order $x_e \sim 10^{-4}$~\cite{Chatterjee:2019jts}. 
\begin{comment}
This plateau reflects the freeze-out of recombination and provides a persistent source of free electrons that keeps the matter weakly coupled to the CMB through Compton scattering until reionization ($z \lesssim 10$), when the first luminous sources begin to ionize the universe once again~\cite{Peebles:1994xt}. 
\end{comment}
With this setup in hand, we express the conversion probability as
\be\label{conv.prob.final}
\mathcal{P}(z, \Delta l)=\frac{g^2_{\phi \gamma}}{4}f({\rm k}_{IR}, {\rm k}_{UV}, B_0, \Delta l)|\mathcal{I}(z,z_0)|^2,
\ee
with 
\be\label{int.conv}
\bea
\mathcal{I}(z,z_0)&=\int_{z_0}^{z}\frac{1}{H_0\sqrt{\Omega_m}}\frac{dz_1}{(1+z_1)^{3/2}}\times\\ &~~~~~~~~~~~~~~~~\exp\left[\frac{iH_0\sqrt{\Omega_m}}{\omega_0\sqrt{1+z_1}}+\frac{i}{H_0\sqrt{\Omega_m}}\left\{\frac{m^2_\phi}{7\omega_0(1+z_1)^{\frac{7}{2}}}-\frac{e^2n_{b0}x_e}{\omega_0m_e(1+z_1)^{\frac{1}{2}}}\right\}\right].
\eea
\ee
Now, we describe below the numerical procedure to evaluate the above expression of conversion probability.

 \subsection{Numerical estimation of the axion-photon conversion probability} 
 \begin{figure}[t]
\centering
\includegraphics[scale=0.5]{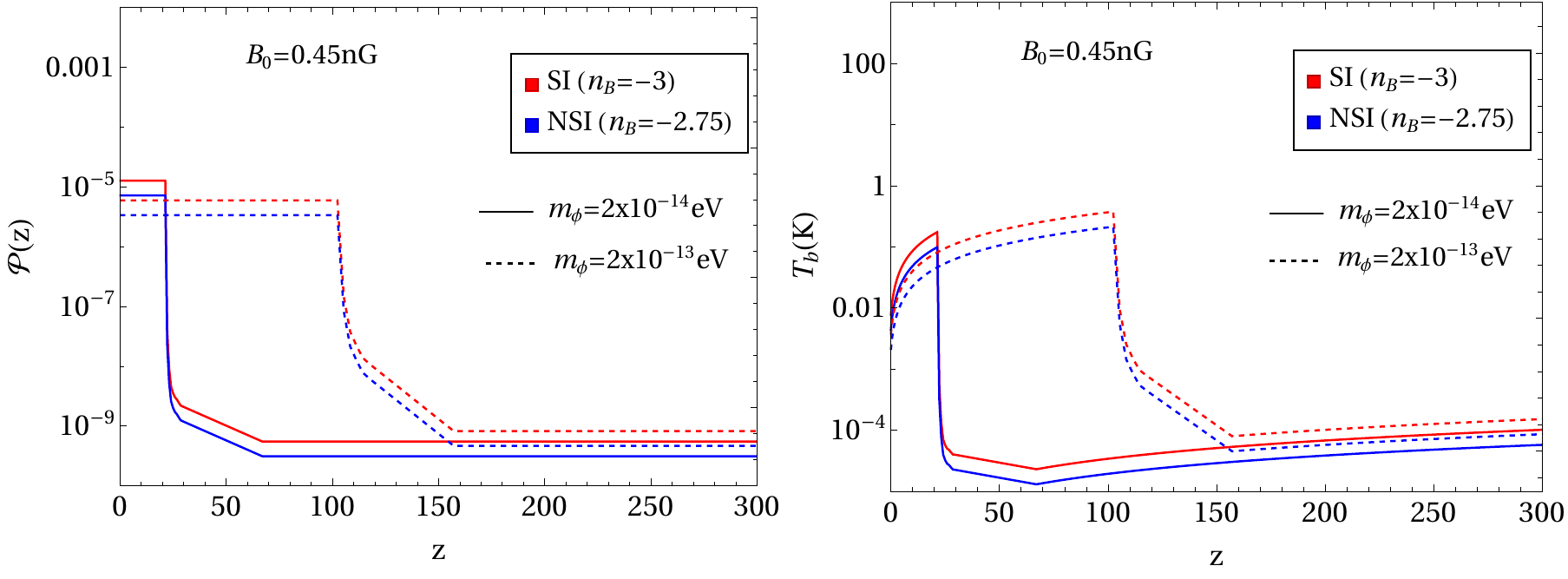}
\caption{{\bf Left panel:} The conversion probability of the axion-photon has been plotted with the redshift factor, z. {\bf Right panel:} The brightness temperature due to the converted photon has been plotted with the redshift factor, z. For both the plot we have considered two different values of the axion mass with same frequency $\omega_0=10 \, {\rm GHz}$. The background magnetic field model has been taken to be scale invariant with the present value, $B_0=0.45 \, {\rm nG}$.}\label{conv_prob_btemp}
\end{figure}
The preceding discussion suggests the occurrence of resonance conversion of axion-photon at a certain redshift determined by the axion mass and the effective photon mass. Therefore, naively integrating the expression for conversion \eqref{int.conv} starting from the recombination to the present may eventually come across stiffness at and around the resonance point. In order to suitably perform this computation, the integration is carried out in a piecewise manner by utilizing the built-in tool NDsolve in \texttt{Mathematica}. The domain of integration is divided into two parts by implementing: $\mathcal{I}(z,z_0)\simeq \mathcal{I}(z,z_m+\epsilon)+\mathcal{I}(z_m-\epsilon,z_0)$ with $z_m$ denoting the resonance point and $\epsilon$ is small parameter $\epsilon(z-z_0)^{-1}<<1$. With this setup, we have been able to obtain a stable numerical result for $\epsilon(z-z_0)^{-1}\sim 10^{-6}$ within the interval $z_0=1100$ and $z=0$. As far as the parameters, [$g_{\phi \gamma}, m_\phi, B_0$], in the current model are concerned,  there exists a strong upper bound on the coupling of ALPs to photon $ g_{\phi \gamma} \leq 6.3 \times 10^{-13}\, \text{GeV}^{-1}$, for most light ALPs of masses $ m_\phi \lesssim 10^{-12}$ eV at 99.7 \% confidence \cite{Sisk_Reyn_s_2021, Conlon_2017}, obtained with X-ray observations of galaxy clusters. On the other hand, given the value of the background magnetic field, $B_0$, constraints from CMB distortion attributes $g_{\phi\gamma}B_0\lesssim 10^{-12} \, {\rm GeV}^{-1} \, {\rm nG}$~\cite{Mirizzi:2009nq}. For the present analysis, considering $m_{\phi} \sim 10^{-14} - 10^{-12} \, \text{eV} $ and $ g_{\phi \gamma} \sim 5 \times 10^{-13} \, \text{GeV}^{-1}$, efficient conversion is found to be possible in the presence of background magnetic fields of the order of ${\rm nG}$. With all the parameters fixed, the remaining variable in evaluating the conversion probability is the frequency, $f=\omega_0/2\pi$, for which the relevant range is $f\in 10~{\rm MHz}-10~{\rm GHz}$ corresponding to ARCADE2 and other low-frequency observations. For this range of frequencies, the conversion probability is evaluated. As a test plot, we have illustrated the behaviour of the conversion probability with respect to the redshift factor in the left plot of Fig.\ref{conv_prob_btemp}. Interestingly, the conversion probability reaches its peak at the resonance point, which, depending upon the axion mass, happens to be at $z\sim 21.6$ for $m_\phi=2\times 10^{-14} \, {\rm eV}$ and at $z\sim 103.8$ for $m_\phi=2\times 10^{-13} \, {\rm eV}$. This is consistent with the earlier discussion on the oscillation length, as illustrated in Fig.~\ref{losc}. Importantly, the photons, thus resonantly converted from axions, have an associated brightness temperature, which could potentially impact the CMB background. This in turn creates the premise to investigate the excess radiation measured by the ARCADE2+Low-frequency observations, and will be our topic of discussion in the next section. 

%%%%%%%%%%%%%%%%%%%%%%%%%%%%%%%%%%%%%%%%%%%%%%%%%
\begin{figure}[t]
\centering
\includegraphics[scale=0.6]{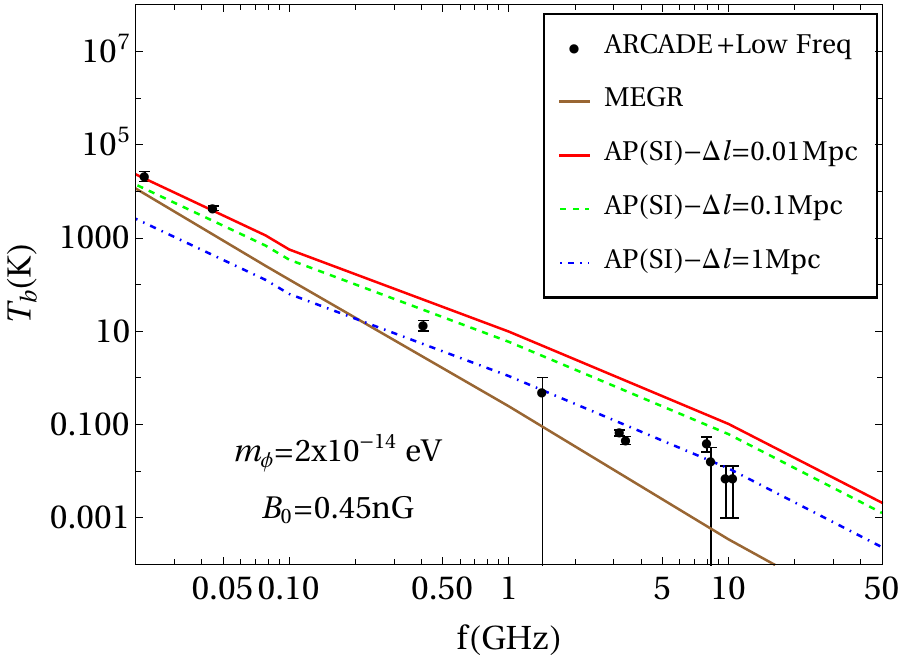}
\caption{In this plot we have presented the excess in the brightness temperature, obtained from the axion-photon conversion, by considering different values for the correlation length, $\Delta l$, of the scale invariant magnetic field.}\label{TbdeltaL}
\end{figure}
%%%%%%%%%%%%%%%%%%%%%%%%%%%%%%%%%%%%%%%%%%%%%%%%%
\begin{figure}[t]
\centering
\includegraphics[scale=0.5]{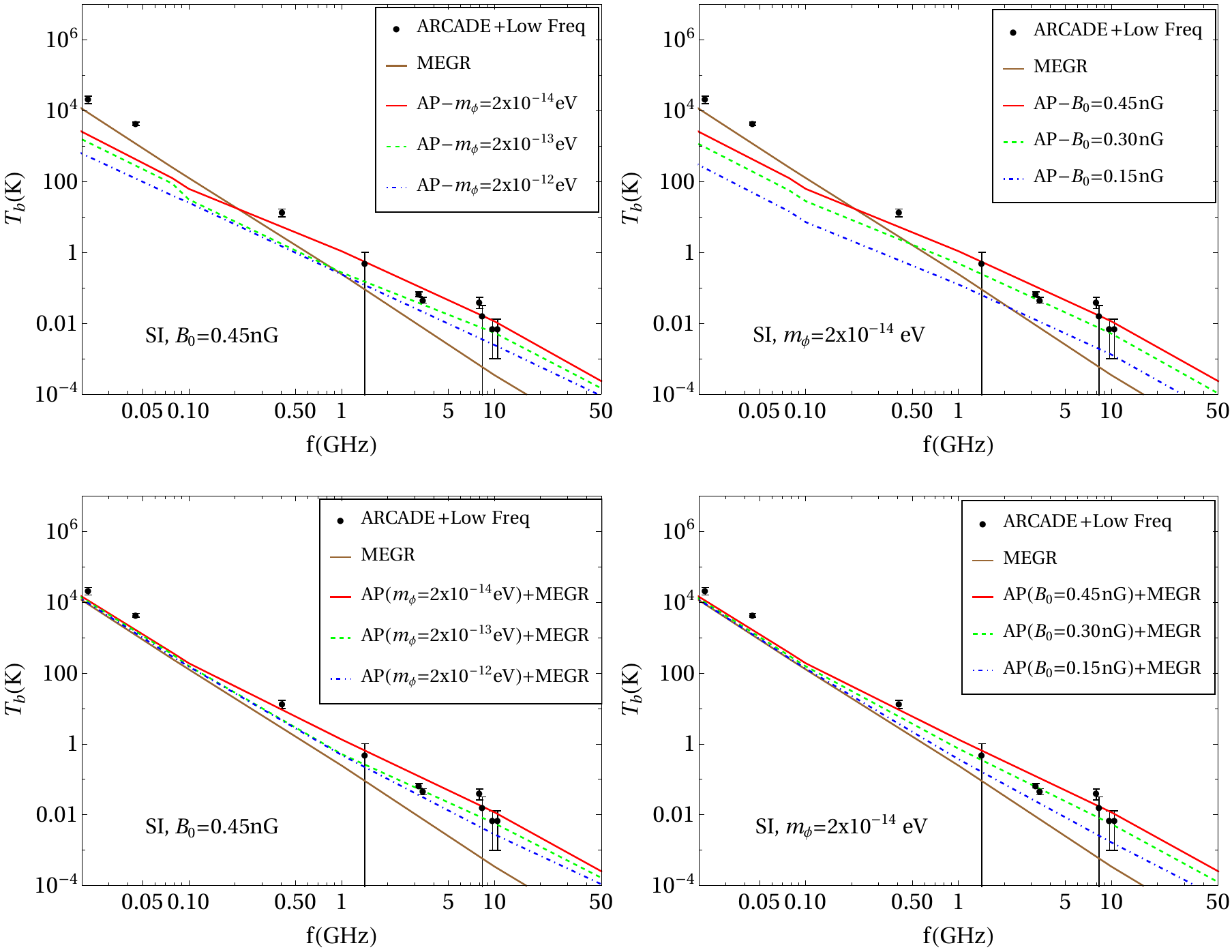}
\caption{The excess in the brightness temperature, obtained from the axion-photon conversion, has been plotted with frequency by considering scale invariant ($n_B=-3$) magnetic power spectrum. {\bf Upper left:} For different axion mass with fixed amplitude of the background magnetic field, {\bf Upper right:} For different amplitude of background magnetic field with fixed axion mass, {\bf Lower left:} We have added the contribution due to the minimum extra galactic radiation (MEGR), for different axion mass, with fixed amplitude of the background magnetic field, {\bf Lower right:} Similar additions have been considering different amplitude of the background magnetic field with fixed axion mass. Notably, we have presented the data of the excess from ARCADE2 and other low-frequency observations in each of the plots. We have also separately shown the line (brown) of the MEGR data for comparison.}\label{Tb_SI}
\end{figure}
%%%%%%%%%%%%%%%%%%%%%%%%%%%%%%%%%%%%%%%%%%
\begin{figure}[t]
\centering
\includegraphics[scale=0.5]{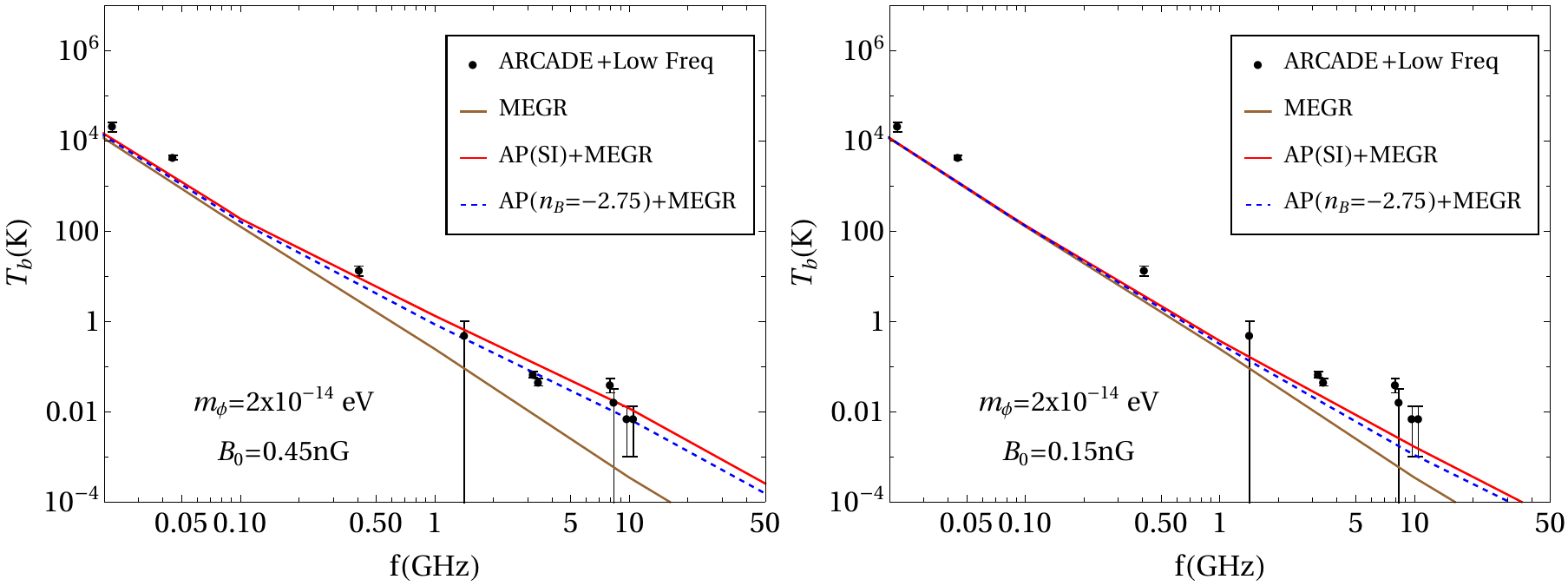}
\caption{In this plot we have presented the excess in the brightness temperature, obtained from the axion-photon conversion, by considering exactly ($n_B=-3$) and nearly scale invariant ($n_B=-2.75$) magnetic power spectra. For the two plots, we have chosen two values of the normalized magnetic field amplitude, {\bf Left panel:} $B_0=0.45 {\rm nG}$, and {\bf Right panel:} $B_0=0.15 {\rm nG}$.}\label{TbdiffnB}
\end{figure}
%%%%%%%%%%%%%%%%%%%%%%%%%%%%%%%%%%%%%%%%%%%%%%%%%
\begin{figure}[t]
\centering
\includegraphics[scale=0.47]{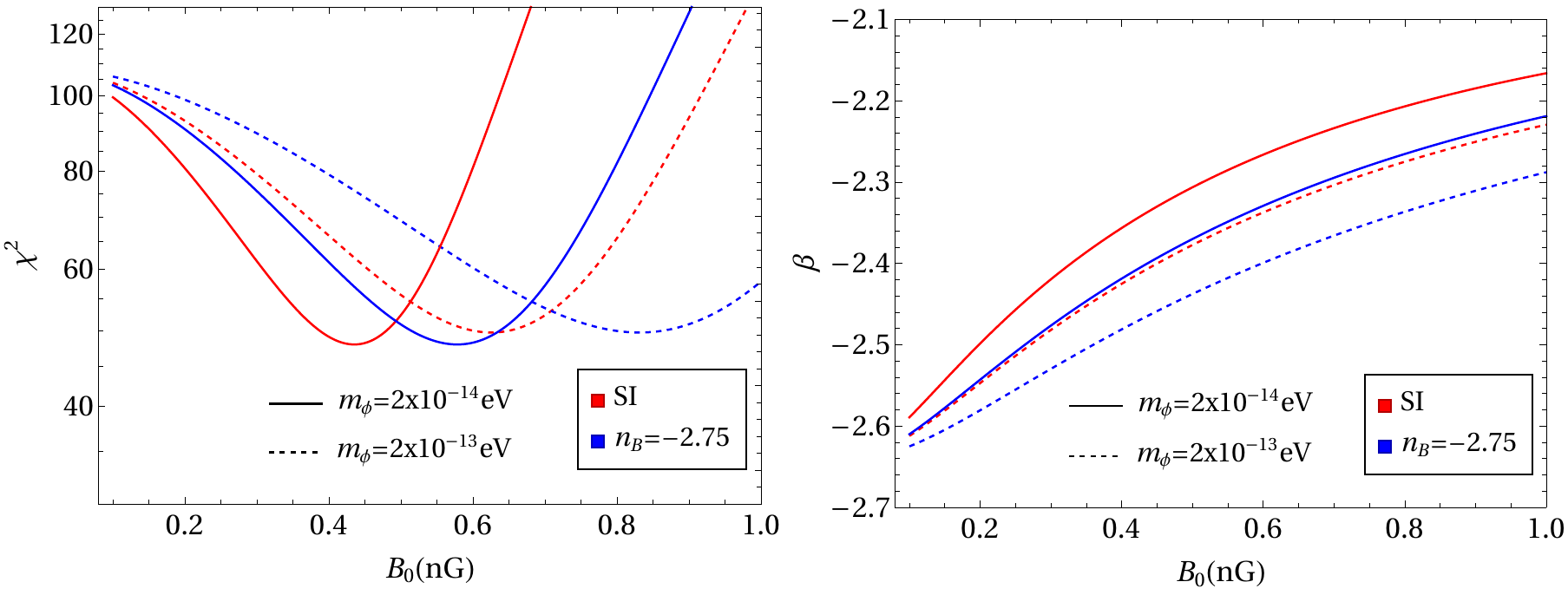}
\caption{{\bf Left panel:} The $\chi^2$ value for fitting the ARCADE2 and low-frequency excess temperature data with our model, which includes contributions from axion–photon conversion and the minimal extragalactic radiation, is plotted with respect to the background magnetic field amplitude. {\bf Right panel:} The exponent, $\beta$, for the same consideration, is plotted as a function of the background magnetic field amplitude. Notably, the power law fit of the measured ARCADE2+Low-frequency data provides for the exponent to be $\beta=-2.6$.}\label{chisqr_exp}
\end{figure}
%%%%%%%%%%%%%%%%%%%%%%%%%%%%%%%%%%%
\section{Axion-photon conversion and brightness temperature}\label{sec-4}
The central observable in our analysis is the brightness temperature \cite{Fixsen2011}, which provides a convenient way to characterize the observed radiation intensity in terms of an equivalent blackbody temperature. To connect the measured photon flux to this quantity, the photon intensity is first related to the underlying spectral distribution, which is governed by Planck’s law,
\be
I(\omega, T)=\frac{\omega^3}{2 \pi^2} \frac{1}{e^{\frac{\omega}{T}} - 1}.
\ee
The associated brightness temperature can be found out analytically with the Rayleigh-Jeans approximation at low frequency limit ($\omega \lesssim 1$ THz), and can be expressed in terms of the frequency as
\be\label{temp.intens}
T_b(\omega, T) = \frac{2 \pi^2}{\omega^2} I(\omega, T).
\ee
In the present context, a fraction of the ALP is converted into photons, which we quantify in terms of the axion-photon conversion probability  as~\cite{Domcke:2020yzq, Ejlli:2019bqj, Fujita:2020rdx},
\be
\Omega_{\gamma}(\omega, T(z)) = \Omega_{\phi}(\omega, T(z)) \mathcal{P}(z, \Delta l),
\ee
where the energy density of the photon per logarithmic frequency interval takes the following form:
\be\label{photon.energy1}
\Omega_{\gamma}(\omega, T(z)) = \frac{1}{\rho_c} \frac{d \rho_{\gamma}(\omega, T(z))}{d(\ln \omega)}=\frac{\omega^4}{\pi^2 \rho_c}\frac{1}{e^{\frac{\omega}{T}} - 1},
\ee
with $\rho_c =3 H_0^2/(8\pi G)$ representing the critical energy density of the universe. Utilizing \eqref{photon.energy1} and \eqref{temp.intens}, the brightness temperature reads,
\be
T_b(\omega, T) = \frac{\pi^4}{15} \frac{T^4}{\omega^3}  \frac{\Omega_{\gamma}(\omega, T)}{\Omega_{\gamma}(T)},
\ee
where, 
\be
\Omega_{\gamma}(T) = \int \Omega_{\gamma}(\omega, T) d(\ln \omega) = \frac{\pi^2 T^4}{15 \rho_c}.
\ee
Bringing in the conversion probability, we arrive at the brightness temperature
\be\label{Tb}
T_b(\omega, T(z)) = \frac{\pi^4}{15} \frac{T^4}{\omega^3} \frac{\Omega_{\phi}(\omega, T)}{\Omega_{\gamma}(T)} \mathcal{P}^{tot}(\omega, z).
\ee
Assuming a frequency-independent ALP abundance, the ALP energy density $\Omega_{\phi}(T) = \int d(\ln \omega) \Omega_{\phi}(\omega, T) \simeq \Omega_{\phi}(\omega, T)$, the above expression for the brightness temperature turns out at present to be
\be
T_b^0(\omega_0) = \frac{\pi^4}{15} \frac{T_0^4}{\omega_0^3} \frac{\Omega_{\phi}^0(\omega_0)}{\Omega_{\gamma}^0} \mathcal{P}^{tot}(\omega_0),
\ee
where we have utilized the following scaling
\be
T = \frac{T_0}{a}, \quad \Omega_{\gamma}(T) = \frac{\Omega_{\gamma}^0}{a^4}, \quad \Omega_{\phi}(T) = \frac{\Omega_{\phi}^0}{a^4}.
\ee
%where the current CMB temperate is $T_0 = 2.725$ K and
Now, the ratio of ALP to photon energy density is defined as
\begin{equation}
\gamma_\phi = \frac{\Omega_{\phi}(\omega, T)}{\Omega_{\gamma}(T)} \simeq \frac{\Omega_{\phi}( T)}{\Omega_{\gamma}(T)} = \frac{\Omega_{\phi}^0}{\Omega_{\gamma}^0},
%= \text{present day ALP-to-photon density ratio}
\end{equation}
where the last term above denote the present day ALP-to-photon density ratio.
This ratio is constrained by $\gamma_\phi = 0.23 \Delta N_{\text{eff}} \lesssim 0.039$ with the extra effective relativistic species $\Delta N_{\text{eff}} \lesssim 0.17$, recently reported in ACT DR6 \cite{ACT:2025tim}.
\begin{table}[h!]
\centering
\begin{tabular}{| l|c|c|r |} % 4 columns: left, center, center, right
\hline
Observation & Frequency & Temperature & Uncertainty \\
& (GHz) & (K) & (K)\\
\hline
Roger & 0.022 & 21200 & 5125 \\
\hline
Maeda & 0.045 & 4355 & 520 \\
\hline
Haslam & 0.408 & 16.24 & 3.4 \\
\hline
Reich & 1.42 & 3.213 & .53 \\
\hline
 & 3.20 & 2.792 & 0.010 \\
\cline{2-4}
 & 3.41 & 2.771 & 0.009 \\
 \cline{2-4}
 & 7.97 & 2.765 & 0.014 \\
 \cline{2-4}
ARCADE 2 & 8.33 & 2.741 & 0.016 \\
 \cline{2-4}
 & 9.72 & 2.732 & 0.006\\
\cline{2-4}
 & 10.49 & 2.732 & 0.006\\
 \cline{2-4}
 & 29.5 & 2.529 & 0.155\\
 \cline{2-4}
 & 31 & 2.573 & 0.076\\
 \cline{2-4}
 & 90 & 2.706 & 0.019\\
\hline
\end{tabular}
\caption{Low-frequency observation and ARCADE 2 data obtained from Ref.~\cite{Fixsen2011}.} 
\label{tab:excessdat}
\end{table}
With these parameters in hand, we evaluate the brightness temperature due to the converted photons, which we argue  to be the cause of the excess temperature observed in ARCADE2+other low-frequency observations. As per our consideration of scale invariant (SI)  magnetic power spectrum as a starting point,  the brightness temperature for such a case has been plotted in Fig.\ref{TbdeltaL}. Importantly, in this figure, we present the results for different correlation length scales, $\Delta l$. Visibly, the brightness temperature tends to saturate for $\Delta l< 1 \, {\rm Mpc}$. Given the coherent length scale of the background magnetic field to be $\sim 1 \, {\rm Mpc}$, $\Delta l$ can be set to be of the same order for our subsequent analysis. Moving forward, the parameter space has been analyzed in terms of the axion mass, $m_\phi$, and the SI magnetic field amplitude, $B_0$, as shown in Fig.\ref{Tb_SI}, for achieving the range of the observed excess data. In this observed spectrum, on the other hand, the minimum extra galactic radiation (MEGR) serves as the baseline contribution from established populations of extragalactic sources, primarily radio galaxies \cite{vernstrom2011contribution} and star-forming galaxies \cite{galluzzi2025updated, ysard2012microwave}. Within the ARCADE2 context, this baseline is of particular importance because it marks the lower limit that conventional astrophysics can explain. However, the measured radio brightness at ARCADE2 and low frequencies significantly exceeds this MEGR level, even when considering a comprehensive set of known radio sources \cite{singal2010sources}. The gap between these two curves is what motivates the exploration of additional mechanisms, such as axion–photon conversion, to account for the unexplained surplus of isotropic radio emission. By simply adding these two contributions, i.e., the converted photons and MEGR, the resulting spectrum aligns remarkably well with the observed radio excess, as illustrated in Fig.\ref{Tb_SI} and in the subsequent figures. Furthermore, in this promising framework, the present model provides a means to examine the structure of the primordial magnetic field, which serves as the background magnetic field in our analysis. In the nearly scale-invariant scenario, variations in the magnetic field lead to observable deviations in the excess spectrum compared to the exactly scale-invariant case, as shown in Fig.\ref{TbdiffnB}. 
\\
Putting the analysis together, we find it convenient to examine the potential parameter space, $\{m_\phi, B_0, n_B\}$, through the implementation of a $\chi^2$ analysis, in which the results are fitted to the observed excess spectrum. The methodology is as follows: a power-law model with two parameters $\mathcal{N}$ and $\omega^{\beta}_0$,  is employed to fit the data obtained from the numerical analysis, with MEGR included. To determine the overall constant factor, $\mathcal{N}$, and the exponent, $\beta$, it is suitable to express them in logarithmic form as $\ln \mathcal{N} + \beta \ln \omega_0$, so that the built-in ``LinearModelFit'' function in \texttt{Mathematica} can be utilized to perform the fitting. We have presented the variation of this exponent, $\beta$, in the right panel of Fig.\ref{chisqr_exp} with respect to the amplitude of the background magnetic field for SI and nearly scale invariant (NSI)   cases of the power spectrum. Once these constants are determined for different choices of parameters, the model is inserted into the definition of $\chi^2 = (y_i - y_{\rm fit}[x_i])^2 / \sigma_i^2$, where $y_i$ represents the observed data from ARCADE2 and low-frequency observations, $y_{\rm fit}$ denotes the corresponding value from the fitted power-law model at each frequency $x_i$, and $\sigma_i$ corresponds to the uncertainty in the observed data. Whereas $i$ signifies the number of data points, which, as per the observations, we have considered $i=10$. Further, we have excluded the last three observed data points from the ARCADE2+Low-frequency measurement, presented in Table.\ref{tab:excessdat} due to their values going below the background CMB temperature $T_0=2.725~{\rm K}$. Using this setup, the $\chi^2$ values have been plotted against the amplitude of the background magnetic field, $B_n$, in the left panel of Fig.\ref{chisqr_exp}, considering both scale-invariant and nearly scale-invariant power spectrum models while also illustrating how the results vary for two different axion masses. It is evident that $\chi^2 \lesssim 100$ for magnetic field amplitudes in the ranges $0.1\,\mathrm{nG} \lesssim B_0 \lesssim 0.6\,\mathrm{nG}$ and $0.1\,\mathrm{nG} \lesssim B_0 \lesssim 0.9\,\mathrm{nG}$ for the SI scenario with axion masses $m_\phi = (2\times 10^{-14},2\times 10^{-13})\,\mathrm{eV}$, respectively. While, for the NSI case, we find a slightly wider parameter range; for example, for $m_\phi = 2\times 10^{-14}\,\mathrm{eV}$ with $\chi^2 \lesssim 100$ corresponds to $0.1\,\mathrm{nG} \lesssim B_0 \lesssim 0.8\,\mathrm{nG}$. In contrast, when only the MEGR is considered, the $\chi^2$ rises to approximately 108, indicating the need for further contributions to the radio excess beyond MEGR alone. Importantly, by including the converted photons, the present model attains a minimum $\chi^2_{\rm min} \sim 48$ for both the SI ($n_B = -3$) and NSI ($n_B = -2.75$) magnetic spectra, corresponding to amplitudes of $0.44\,\mathrm{nG}$ and $0.58\,\mathrm{nG}$, respectively, assuming $m_\phi \sim 2 \times 10^{-14}\,\mathrm{eV}$. The associated power-law slopes are $\beta \sim -2.34$, while the broader parameter space with $\chi^2 \lesssim 100$ corresponds to $\beta \sim [-2.6, -2.3]$.\\
Having established that axion–photon conversion can reproduce the observed excess in the radio brightness temperature, it is natural to examine its implications for the global 21-cm signal. Since the depth of the absorption trough is highly sensitive to the background radiation field as well as the thermal history of the intergalactic medium, the same mechanism that explains ARCADE2 could also impact the EDGES anomaly. We therefore turn to this connection in the next section.

%%%%%%%%%%%%%%%%%%%%%%%%%%%%%%%%%%%%%%%%%%%
\section{Impact on the global 21-cm signal (EDGES)}\label{sec-5}
\begin{figure}[ht]
    \centering
    \begin{minipage}{0.48\textwidth}
        \includegraphics[width=\linewidth]{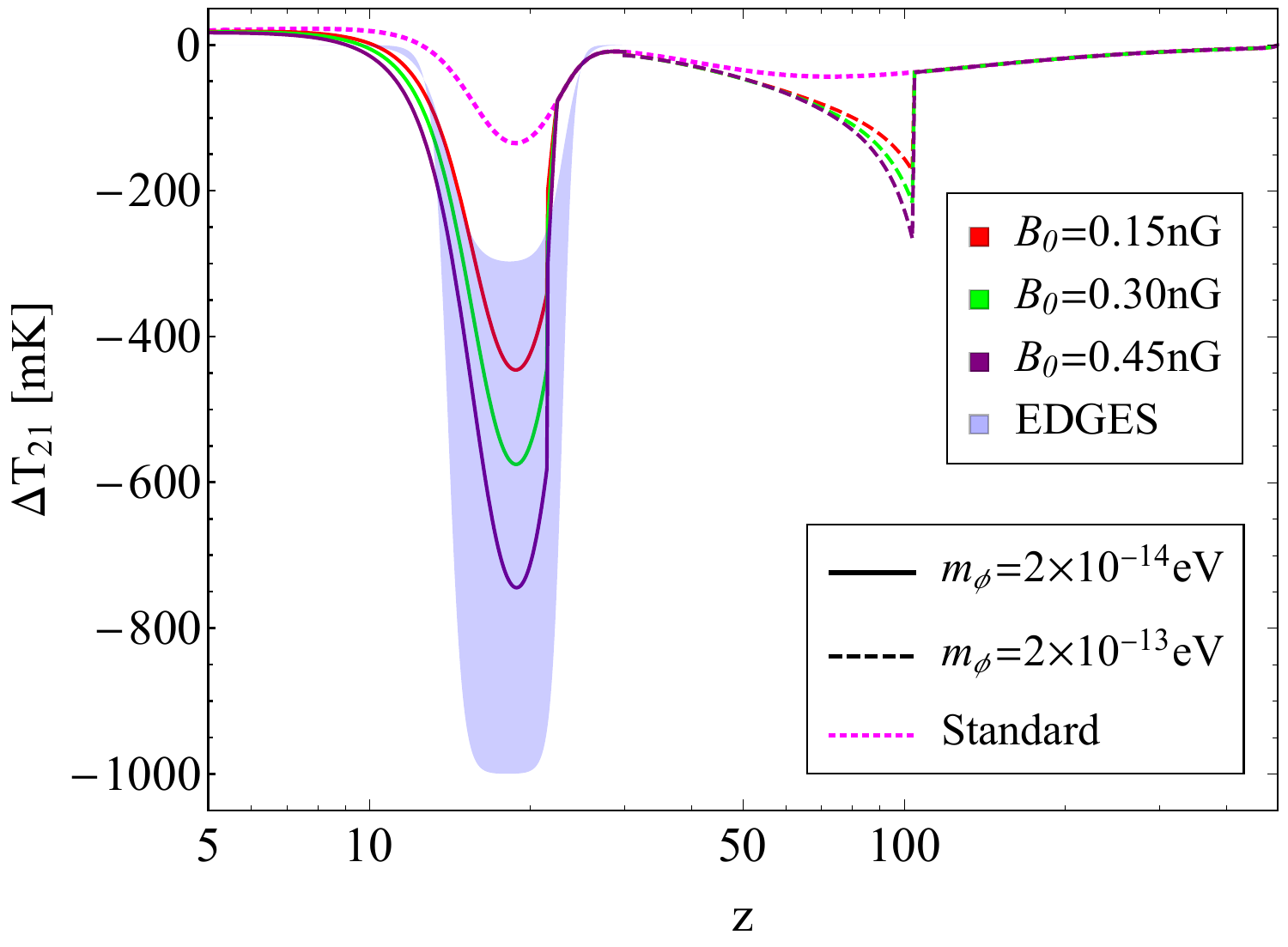}
    \end{minipage}
    \hfill
    \begin{minipage}{0.48\textwidth}
        \includegraphics[width=\linewidth]{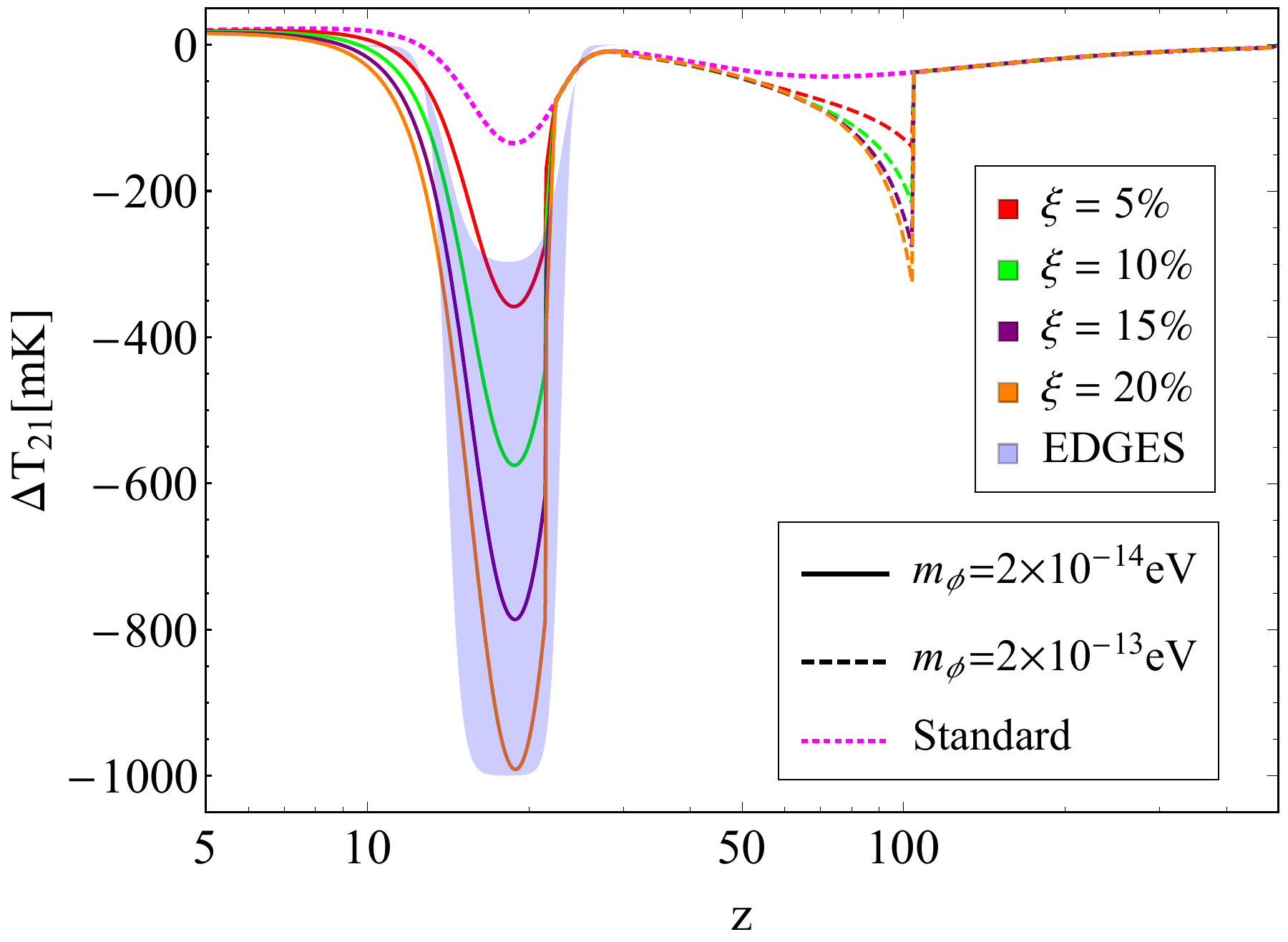}
    \end{minipage}
    \caption{The global 21-cm brightness temperature contrast relative to the background radiation for two benchmark masses with scale invariant $(n_B = -3)$ magnetic power spectrum. The solid curves ($m_\phi = 2 \times 10^{-14}$ eV) indicate an enhanced absorption trough compared to the standard prediction (magenta dotted), consistent with the EDGES band (blue) during cosmic dawn. The benchmark with $m_\phi = 2 \times 10^{-13}$ eV exhibits a distinctive kinematic feature, showing an additional absorption trough (dashed) during dark ages. Results are shown for different background magnetic fields with fixed $\xi = 10\%$ (left panel) and for different fractional contributions with fixed $B_0 = 0.3$nG. (right panel).}\label{21cm}
\end{figure}

The principal observable in 21-cm cosmology is the \textit{differential brightness temperature} $\Delta T_{21}$, which quantifies the contrast between the spin temperature of neutral hydrogen and the background radiation temperature. In the absence of astrophysical foregrounds, the sky-averaged (global) differential brightness temperature is given by~\cite{Furlanetto2006, Pritchard2012, Madau1997}  
\begin{equation}
\Delta T_{21}(z) \simeq 27\, x_{\mathrm{HI}} \left( \frac{\Omega_{b,0} h^2}{0.023} \right) \sqrt{ \frac{0.15}{\Omega_{m,0} h^2} \frac{1+z}{10} } \left(1 - \frac{T_R}{T_s} \right) \, \mathrm{mK},
\label{eq:T21}
\end{equation}  
where $x_{\mathrm{HI}}$ is the neutral hydrogen fraction, $T_R(z)$ is the total radio background at redshift $z$, and $T_s(z)$ is the spin temperature of neutral hydrogen (see Appendix~\ref{Global} for details). Extra photons generated through axion--photon conversion contribute an additional background component. At redshift $z$, the corresponding brightness temperature (\ref{Tb}) can be expressed as  
\begin{equation}
T_b(z) = \frac{\pi^4}{15} \, \gamma_\phi \left( \frac{T_0}{\tilde{\omega}} \right)^3 \mathcal{P}^{\rm tot}(\tilde{\omega}, z) \, T_0 \, (1 + z),
\end{equation}  
where $\tilde{\omega} = \frac{2\pi f_{21\mathrm{cm}}}{1+z}$ with $f_{21\mathrm{cm}} = 1.4~\mathrm{GHz}$, and we set $\gamma_\phi = 0.03$ in this work. Further, the total radio background temperature is written as  
\begin{equation}
T_R(z) = T_0 (1 + z) + \xi \, T_b(z),
\end{equation}  
where $\xi$ is a free parameter controlling the fractional contribution of the axion-induced excess radiation to the intergalactic medium~\cite{Feng:2018rje}. \\
\subsection{Scale-invariant scenario}
In the absence of primordial magnetic field heating, as in the case of a scale-invariant PMF spectrum, the baryon kinetic temperature cools adiabatically after thermal decoupling, scaling as $T_k \propto (1+z)^2$.  In the absence of astrophysical heating, collisional coupling ensures $T_s \simeq T_k < T_R$, producing a clean absorption trough in $\Delta T_{21}$ during dark ages. After the redshift $z \sim 30$, the Lyman-$\alpha$ photons generated by the earliest stars efficiently couple the spin temperature to the gas temperature via the Wouthuysen–Field effect, $T_s \simeq T_k$ \cite{1952AJ.....57R..31W, Field1958}. Since $T_k \ll T_R$ during cosmic dawn, resonant photon injection at $z_{\rm res} \sim 20$, corresponding to a benchmark axion mass $m_\phi \sim 2 \times 10^{-14}\,\mathrm{eV}$, drives the contrast factor $(1 - T_R/T_s)$ strongly negative. This produces a deep absorption trough in $\Delta T_{21}$, consistent with the anomalous deep feature reported by EDGES~\cite{Bowman:2018yin}.\\
In the left panel of Fig.~\ref{21cm}, we illustrate this behavior by fixing the excess radiation fraction at $\xi = 10\%$. For representative PMF strengths $B_0 = 0.15, 0.3,$ and $0.45\,\mathrm{nG}$, the predicted absorption depth lies within the EDGES-allowed region $-1 \, \mathrm{K} \lesssim \Delta T_{21} \lesssim -0.3 \, \mathrm{K}$. The right panel of Fig.~\ref{21cm} explores the variation with $\xi$. If the full radio excess ($\xi = 100\%$) is assumed to contribute, the absorption overshoots the EDGES band. To remain consistent with the data, the axion-induced component must be restricted to $\xi \sim 5\%\!-\!20\%$, which keeps the trough depth within the observed range of EDGES anomalies~\cite{Caputo:2020avy}. We have additionally demonstrated how resonant photon injection results in distinctive spectral features in the brightness temperature during the dark ages around $z \sim 100$, corresponding to the benchmark with $m_\phi = 2 \times 10^{-13}$ eV [see Fig.~\ref{21cm}]. Such a characteristic trough would lie well within the reach of next-generation 21-cm experiments and could be readily distinguished from standard astrophysical backgrounds.\\
Taken together, these results show that axion-photon conversion in the presence of nanogauss-level, scale-invariant primordial magnetic fields can naturally account for both the ARCADE2 excess and the EDGES anomaly. However, when the PMF spectrum deviates slightly from scale invariance, magnetic dissipation through ambipolar diffusion and turbulent decay injects heat into the intergalactic medium, thereby suppressing the absorption signal. We now turn to this scenario, where PMF heating becomes important and the allowed parameter space is correspondingly modified. 

\subsection{Nearly scale-invariant scenario: Effect of PMF heating}
\begin{figure}[ht]
    \centering
    \begin{minipage}{0.48\textwidth}
        \includegraphics[width=\linewidth]{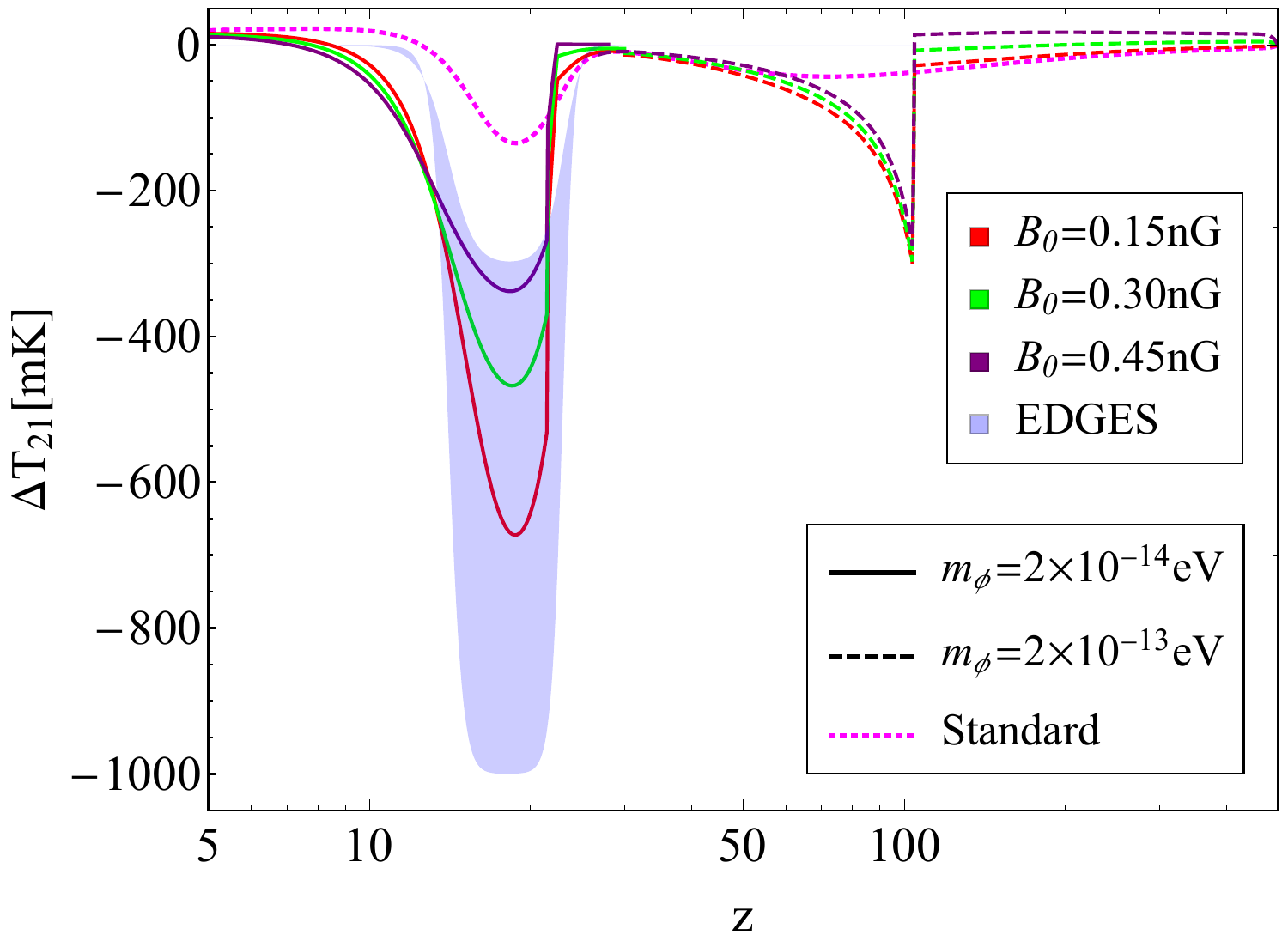}
    \end{minipage}
    \hfill
    \begin{minipage}{0.48\textwidth}
        \includegraphics[width=\linewidth]{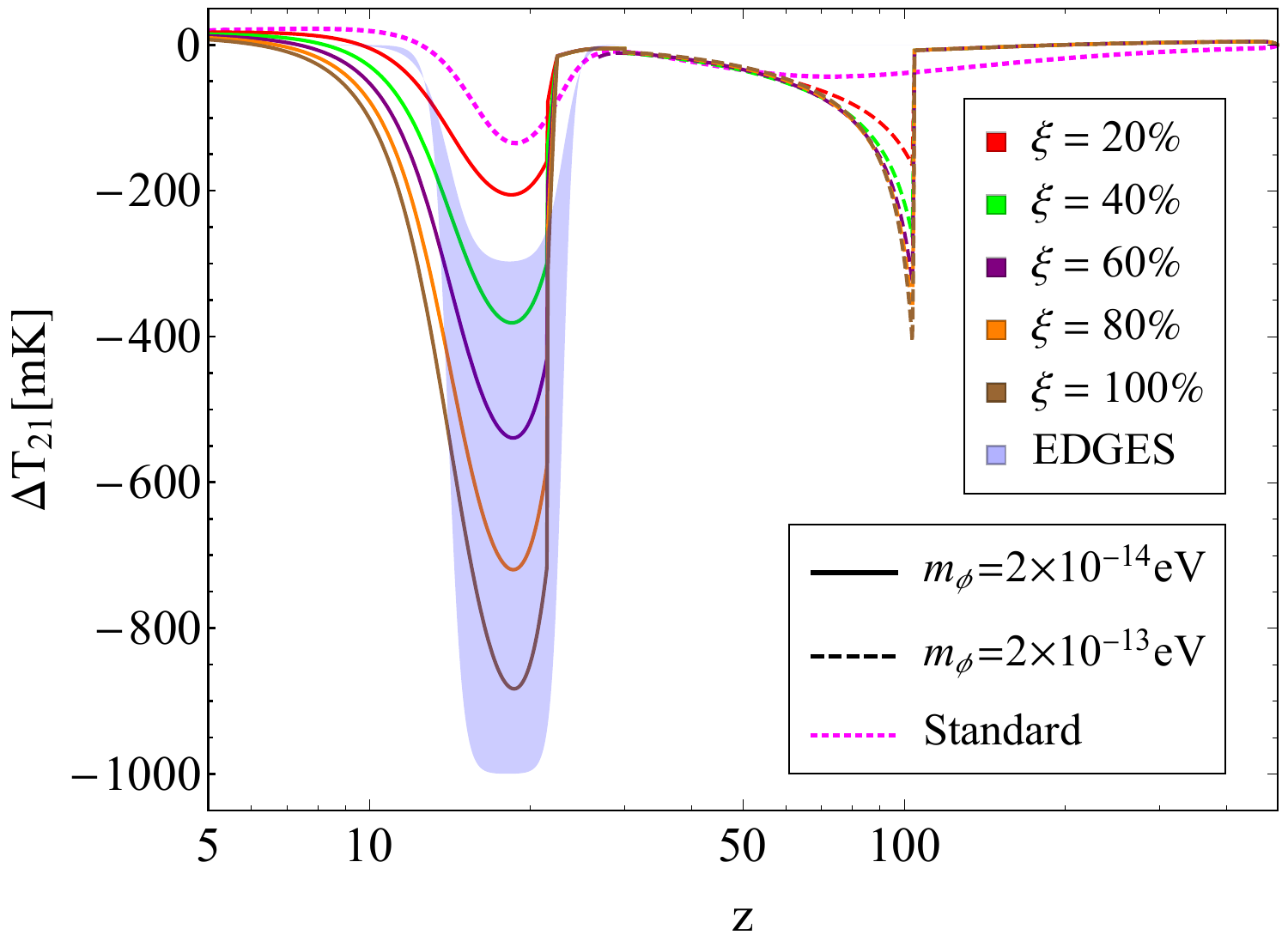}
    \end{minipage}
    \caption{The 21-cm global signal under the excess radio background for two benchmark masses with nearly scale invariant $(n_B = -2.75)$ magnetic power spectrum. Results are shown for different background magnetic fields with fixed $\xi = 50\%$ (left panel) and for different fractional contributions with fixed $B_0 = 0.3 \rm{nG}$ (right panel), compared with EDGES allowed region (blue band) during cosmic dawn.}\label{21cmPMF}
\end{figure}
For the scale-invariant case, no significant heating arises from PMF dissipation (see Appendix~\ref{Global}). However, when the magnetic spectrum deviates slightly from scale invariance (e.g., $n_B = -2.75$), PMFs inject additional energy into the baryons through ambipolar diffusion and decaying turbulence. This raises the gas temperature above its adiabatic value, $T_k(z) = T_k^{\rm ad}(z) + \Delta T_k^{\rm PMF}(z)$, where $\Delta T_k^{\rm PMF}$ encodes the cumulative heating from PMF dissipation. Since the spin temperature $T_s$ is strongly coupled to the gas during cosmic dawn, this heating increases $T_s$ and thereby reduces the contrast with the background radiation temperature $T_R$. As a result, the absorption trough in $\Delta T_{21}$ becomes shallower compared to the scale-invariant case without heating [see Fig.~\ref{21cmPMF}, left panel].\\
In extreme cases of strong magnetic dissipation, $T_s$ can even exceed $T_R$, erasing the absorption feature or transforming it into emission. Thus, the strength of PMF heating directly regulates the depth of the trough: the stronger the heating, the shallower the feature. In our analysis of the nearly scale-invariant case, we varied the fractional contribution of the excess radiation from $\xi = 20\%$ to $100\%$. As shown in the right panel of Fig.~\ref{21cmPMF}, consistency with the EDGES band requires $\xi$ in the range $40\%\!-\!100\%$. This contrasts with the scale-invariant case, where only $\xi \sim 5\%\!-\!20\%$ is allowed. \\
Together, these results highlight the competing roles of resonant photon injection, which deepens the absorption feature, and PMF heating, which counteracts it. The interplay between these two effects defines a narrow but viable region in parameter space where the EDGES anomaly can be accommodated without violating cosmological bounds.  \\
\section{Conclusion and outlook}\label{sec-6}
In this work, we have explored axion-photon conversion in an expanding FLRW spacetime as a unified origin of the ARCADE2 radio excess and the EDGES 21-cm anomaly. Starting from the action, we derived the full equations of motion, including the effect of expanding background and plasma effects, and computed the resonant conversion probability in the presence of stochastic primordial magnetic fields. We demonstrated that axion-like particles with masses in the range $m_\phi \sim 10^{-14} - 10^{-12}\,\mathrm{eV}$, coupled to nanogauss-strength magnetic fields with spectral indices close to scale invariance, can generate soft photons in the MHz-GHz range sufficient to explain the observed isotropic radio excess. The resulting excess brightness temperature reproduces the ARCADE2+low-frequency data well (Fig.~\ref{Tb_SI}, \ref{TbdiffnB}). In particular, our fits yield $\chi^2_{\rm min}\sim 48$ for scale-invariant ($n_B=-3$) and nearly scale-invariant ($n_B=-2.75$) magnetic spectra with amplitudes $0.44\,\mathrm{nG}$ and $0.58\,\mathrm{nG}$, respectively, assuming $m_\phi \sim 2\times 10^{-14}\,\mathrm{eV}$. The corresponding power-law slopes are $\beta \sim -2.34$, while the broader parameter space with $\chi^2 \lesssim 100$ yields $\beta \sim [-2.6,-2.3]$, which we use to investigate the EDGES anomaly. \\
This parameter space enhances the global 21-cm absorption, producing a trough consistent with the EDGES band (Fig.~\ref{21cm}). For a scale-invariant magnetic spectrum, $m_\phi \sim 2\times 10^{-14}\,\mathrm{eV}$ with $B_0 \simeq 0.1\!-\!0.5\,\mathrm{nG}$ yields an absorption depth variation of $\Delta T_{21} \sim -(300\!-\!700)\,\mathrm{mK}$ around $z\simeq 20$, assuming an excess radiation fraction of $\xi=10\%$. A representative benchmark, $m_\phi \sim 2\times 10^{-14}\,\mathrm{eV}$, $B_0 \sim 0.3\,\mathrm{nG}$, and $f_{*}=3\%$, reproduces the EDGES anomaly when $\xi$ is restricted to $\sim 5\%\!-\!20\%$ [see Fig.~\ref{21cm}, right panel]. For larger ALP masses, an additional absorption feature is predicted during the dark ages, providing a distinctive, astrophysics-independent probe of non-standard physics.\\
At the same time, nearly scale-invariant spectra highlight the competing role of primordial magnetic field heating, which suppresses the trough depth, while strictly scale-invariant spectra predict minimal heating. PMFs dissipate energy into the intergalactic medium through ambipolar diffusion and turbulent decay after recombination (see Appendix~\ref{Global}), thereby shifting the allowed parameter space. Nevertheless, consistency with EDGES can still be maintained for $\xi \sim 40\%\!-\!100\%$ with the same benchmark parameters, $m_\phi \sim 2\times 10^{-14}\,\mathrm{eV}$, $B_0 \sim 0.3\,\mathrm{nG}$, and $f_{*}=3\%$ [see Fig.~\ref{21cmPMF}, right panel]. Importantly, the scenario remains consistent with current limits from CMB spectral distortions and $\Delta N_{\mathrm{eff}}$.\\
Taken together, our results demonstrate that global 21-cm observations offer a powerful probe of the ARCADE2 excess and the EDGES anomaly while simultaneously constraining the properties of ALPs and PMFs, thereby opening a unique window into dark-sector physics and the origin of cosmic magnetism. Our analysis further identifies a narrow but well-motivated region in the ALP–magnetic field parameter space where resonant axion–photon conversion can account for both anomalies. \\
Looking ahead, the upcoming 21-cm experiments and radio surveys such as LEDA, PRIZM, REACH~\cite{Price2018LEDA, Chiang:2020pbx, deLeraAcedo:2022kiu}, together with next-generation CMB missions such as PIXIE~\cite{Kogut:2024vbi} and PRISM~\cite{PRISM:2013fvg}, are expected to improve these constraints further.   In particular, joint measurements of spectral distortions, low-frequency radio backgrounds, and the global 21-cm signal can offer critical tests of this framework and could help clarify whether axion–photon conversion plays a role in the observed low-frequency cosmic anomalies. A joint forecast analysis with mock data from such upcoming missions in combination with the exsisting data from current observations, thereby investigating for the synergy of different missions in a wide frequency range, would definitely be an important addition to the present work. We look forward to engaging into some of these works in near future.
%%%%%%%%%%%%%%%%%%%%%%%%%%%%%%%%

\appendix
\addcontentsline{toc}{section}{Appendices}
\section*{Appendices}

\section{Derivation of the axion-photon conversion probability}\label{derive.prob}
The dynamics of ALP-photon mixing in an expanding spacetime with a large-scale magnetic field background has been discussed in the main text, and a matrix equation has been presented in \eqref{meq}. Here, we outline the procedure to derive the conversion probability from this step. We express the matrix equation in the following manner,
\be\label{evo_eq}
\pr_\eta \begin{pmatrix}
A^x\\
A^y\\
\Tilde{\phi}
\end{pmatrix}= \mi \mathcal{K}
    \begin{pmatrix}
A^x\\
A^y\\
\Tilde{\phi}
\end{pmatrix} \ ,
\ee
where $\mathcal{K}$ contains the mixing elements. To simplify our analysis, we expand the mixing matrix perturbatively (in the order of the coupling constant) as $\mathcal{K}=\mathcal{K}_0+\delta\mathcal{K}(\eta)$, where
\begin{comment}
\be
\mathcal{K} = \begin{pmatrix}
        \omega+\frac{\omega^2_{pl} a^2(\eta)}{2\omega} & 0 & \frac{1}{2} g_{\phi \gamma} a^2(\eta) B^x_{0} (1 + \frac{\mi \mathcal{H}(\eta)}{\omega}) \\
        0 & \omega+\frac{\omega^2_{pl} a^2(\eta)}{2\omega} & \frac{1}{2} g_{\phi \gamma} a^2(\eta) B^y_{0}\left(1 + \frac{\mi \mathcal{H}(\eta)}{\omega}\right) \\
        -\frac{i}{2} g_{\phi \gamma}  B^x_{0} (1 + \frac{\mi \mathcal{H}(\eta)}{\omega}) & -\frac{i}{2} g_{\phi \gamma} B^y_{0} (1 + \frac{ \mi \mathcal{H}(\eta)}{\omega}) & \omega+\frac{m^2_\phi a^2(\eta)}{2\omega}-\mi\mathcal{H}(\eta)
        \end{pmatrix}
\ee
\end{comment}
\be\label{K0}
\mathcal{K}_0 = \begin{pmatrix}
        \omega + \Delta_{pl} & 0 & 0 \\
        0 & \omega + \Delta_{pl} & 0 \\
        0 & 0 & \omega + \Delta_\phi
    \end{pmatrix}, 
\ee
with $\Delta_\phi = \frac{m_{\phi}^2 a^2(\eta)}{2\omega} $ and $\Delta_{pl} = \frac{\omega^2_{pl} a^2(\eta)}{2\omega}$, represents the zeroth order matrix. Whereas the perturbation around the background is given by,
\be\label{delK}
\delta \mathcal{K}(\eta) = \begin{pmatrix}
        0 & 0 & \frac{1}{2} g_{\phi \gamma} a^2(\eta) B^x_{0} (1 + \frac{\mi \mathcal{H}(\eta)}{\omega}) \\
        0 & 0 & \frac{1}{2} g_{\phi \gamma} a^2(\eta) B^y_{0} (1 + \frac{\mi \mathcal{H}(\eta)}{\omega}) \\
        -\frac{i}{2} g_{\phi \gamma}  B^x_{0} (1 + \frac{\mi \mathcal{H}(\eta)}{\omega}) & -\frac{i}{2} g_{\phi \gamma} B^y_{0} (1 + \frac{\mi \mathcal{H}(\eta)}{\omega}) &  -\mi \mathcal{H}(\eta)
    \end{pmatrix}.
\ee
Due to this construction, the evolution equation \eqref{evo_eq}, now, simply reads,
\begin{comment}
\footnote{From now on, we take the fields as a function of $\eta$ only, since the weakly dependence on $l$ comes only through the background magnetic field $B_0$. Later we will take care of it when the power spectrum will be calculated.}
\end{comment}
\begin{equation}
    \partial_{\eta} \textbf{A}(\eta,l) = \mi (\mathcal{K}_0(\eta) + \delta \mathcal{K}(\eta,l) ) \textbf{A}(\eta,l) \ \, , \, \ \textbf{A}(\eta,l) = \begin{pmatrix}
        A^x(\eta,l) \\
        A^y(\eta,l) \\
        \Tilde{\phi}(\eta,l)
    \end{pmatrix}.
\end{equation}
This equation can exactly be solved by considering a unitary mixing matrix $\mathcal{U}$ such as~\cite{Addazi:2024mii}
\be\label{unitary}
\bea
    \textbf{A}(\eta, l) &= \mathcal{U}(\eta, \eta_0,l) \textbf{A}(\eta_0, l),\\
    \partial_{\eta} \mathcal{U}(\eta, \eta_0,l) &= \mi (\mathcal{K}_0(\eta) + \delta \mathcal{K}(\eta,l) ) \mathcal{U}(\eta, \eta_0,l).
\eea
\ee
We see that, in the case $g_{\phi \gamma} \to 0$ and $\omega \gg  \mathcal{H}(\eta)$, $\delta \mathcal{K} \to 0$ gives the solution in terms of an instantaneous unitary evolution, 
\begin{equation}\label{U0}
    \textbf{A}_0(\eta, l) = \mathcal{U}_0(\eta, \eta_0) \textbf{A}_0(\eta_0,l) \implies \mathcal{U}_0(\eta, \eta_0) = e^{\mi \int_{\eta_0}^{\eta} d\eta_1 \mathcal{K}_0(\eta_1)}.
\end{equation}
Notably, the zeroth-order unitary operator $\mathcal{U}_0(\eta,\eta_0)$ is independent of $l$, since $l$ enters only through the magnetic field, which contributes at first order. Then the perturbative solution of \eqref{unitary} is given by
\begin{equation}\label{conv}
    \mathcal{U}(\eta, \eta_0,l) = \mathcal{U}_0(\eta, \eta_0) + \mi \mathcal{U}_0(\eta, \eta_0) \int_{\eta_0}^{\eta} d\eta_1 \mathcal{U}^{\dagger}_0(\eta_1, \eta_0) \delta \mathcal{K}(\eta_1,l) \mathcal{U}_0(\eta_1, \eta_0)  + \mathcal{O}(\delta \mathcal{K}^2).
\end{equation}
From axion to photon production at time $\eta$ can be read from the conversion matrix \ref{conv} such as
\be
\bea
    A^x(\eta,l) &= \mathcal{U}_{13}(\eta, \eta_0,l) \Tilde{\phi}(\eta_0,l),\\
    A^y(\eta,l) &= \mathcal{U}_{23}(\eta, \eta_0,l) \Tilde{\phi}(\eta_0, l).
\eea
\ee
Thus, the conversion probability can be defined as
\begin{equation}
    \mathcal{P}(\eta,l_1,l_2) =\mathcal{U}_{13}(\eta, \eta_0,l_1)\mathcal{U}^*_{13}(\eta, \eta_0,l_2)+ \mathcal{U}_{23}(\eta, \eta_0,l_1)\mathcal{U}^*_{23}(\eta, \eta_0,l_2),
\end{equation}
where * indicates the complex conjugation. To keep our analysis on a general footing, we introduce two separate spatial positions along the $l$ axis. We interpret this as the probability of conversion for the ALP-photon system, emerging from the correlation between the two spatial points, which evolves with time starting from $\eta_0$.
%, which can be conveniently done given the fact that we are not including the evolution along  
With the help of \eqref{K0}, \eqref{delK}, \eqref{U0}, and \eqref{conv}, and performing a straightforward algebraic manipulation, we obtain the following two-point correlators of the unitary matrix element, 
\be
\bea
|\mathcal{U}_{13}|^2 = &\frac{g_{\phi \gamma}^2}{4}
\int_{\eta_0}^{\eta} d\eta_1 \int_{\eta_0}^{\eta} d\eta_2 a^2(\eta_1) B^x_0(\eta_1, l_1) a^2(\eta_2) B^x_0(\eta_2, l_2) \left(1 + \frac{\mi \mathcal{H}(\eta_1)}{\omega_1} \right)
\left(1 - \frac{\mi \mathcal{H}(\eta_2)}{\omega_2} \right)\\
&e^{-\mi \int_{\eta_1}^{\eta_2} d\eta_3 [\Delta_{pl}(\eta_3) - \Delta_\phi(\eta_3) ]}, \\
|\mathcal{U}_{23}|^2 = &\frac{g_{\phi \gamma}^2}{4}
\int_{\eta_0}^{\eta} d\eta_1 \int_{\eta_0}^{\eta} d\eta_2 a^2(\eta_1) B^y_0(\eta_1, l_1) a^2(\eta_2) B^y_0(\eta_2, l_2) \left(1 + \frac{\mi \mathcal{H}(\eta_1)}{\omega_1} \right)
\left(1 - \frac{\mi \mathcal{H}(\eta_2)}{\omega_2} \right) \\
&e^{-\mi \int_{\eta_1}^{\eta_2} d\eta_3  [\Delta_{pl}(\eta_3) - \Delta_\phi(\eta_3) ]}.
\eea
\ee
To proceed, we find it convenient to express the background magnetic field in terms of its value at the present epoch of the universe. For a large-scale isotropic and homogeneous background magnetic field, we obtain $a^2(\eta) B_0(\eta, l)\sim \tilde{B}_0(l)$. With this setup, the conversion probability reads,
\be
\bea
\mathcal{P}(\eta,l_1,l_2) = &\frac{g_{\phi \gamma}^2}{4}
\int_{\eta_0}^{\eta} d\eta_1 \int_{\eta_0}^{\eta} d\eta_2 [ \Tilde{B}^x_0(l_1) \Tilde{B}^x_0(l_2) + \Tilde{B}^y_0(l_1) \Tilde{B}^y_0(l_2) ] \left(1 + \frac{\mi \mathcal{H}(\eta_1)}{\omega_1} \right) \left(1 - \frac{\mi \mathcal{H}(\eta_2)}{\omega_2} \right) \\
&e^{-\mi \int_{\eta_1}^{\eta_2} d\eta_3  [\Delta_{pl}(\eta_3) - \Delta_\phi(\eta_3) ]}.
\eea
\ee
For a nearly isotropic Gaussian distribution, the two-point correlation function of the magnetic field is usually expressed as \cite{Kobayashi_2019},
\be
\langle \tilde{B}^x_0(l_1) \tilde{B}^x_0(l_2) + \tilde{B}^y_0(l_1) \tilde{B}^y_0(l_2) \rangle = \int \frac{d^3{\rm k}}{(2\pi)^3} e^{\mi{\rm k}\cos \theta (l_1 - l_2)} P_B({\rm k}).
\ee
Finally, the conversion probability in the integral form
\be
\bea
\mathcal{P}(\eta,\Delta l) = &\frac{g_{\phi \gamma}^2}{4}
\int_{\eta_0}^{\eta} d\eta_1 \int_{\eta_0}^{\eta} d\eta_2 \int \frac{d^3 \rm{k}}{(2\pi)^3} e^{\mi \rm{k} \Delta l \cos \theta} P_B(\rm k) \left(1 + \frac{\mi \mathcal{H}(\eta_1)}{\omega_1} - \frac{\mi \mathcal{H}(\eta_2)}{\omega_2} \right) \\
&e^{-\mi \int_{\eta_1}^{\eta_2} d\eta_3  [\Delta_{pl}(\eta_3) - \Delta_\phi(\eta_3) ]},
\eea
\ee
where $\Delta l=l_1-l_2$.
\section{Global 21-cm signal and impact of PMF on IGM temperature}\label{Global}
The spin temperature $T_s$, which determines the relative population of the hydrogen hyperfine levels, is influenced by three primary mechanisms: radiative coupling to the CMB, collisional coupling with the gas, and resonant scattering of Lyman-$\alpha$ photons (Wouthuysen–Field effect). The spin temperature $T_s$ is given by a weighted harmonic mean~\cite{Furlanetto2006, Pritchard2012}
\begin{equation}
T_s^{-1} = \frac{T_R^{-1} + x_c T_k^{-1} + x_\alpha T_\alpha^{-1}}{1 + x_c + x_\alpha}.
\label{eq:Ts}
\end{equation}
Here, the coefficients $x_c$ and $x_\alpha$ quantify the efficiency of collisional and Ly$\alpha$ coupling, respectively, $T_k$ is the gas kinetic temperature, and $T_\alpha$ is the color temperature of the Lyman-$\alpha$ field. Note that the differential brightness temperature $\Delta T_{21}$ \eqref{eq:T21} saturates when the spin temperature greatly exceeds the background radiation temperature, $T_s \gg T_R$, leading to a maximum emission signal~\cite{Pritchard2012}. In contrast, when $T_s \ll T_R$, the signal can become arbitrarily large in absorption, as the contrast term $ 1 - T_R/T_s$ becomes significantly negative. In the absence of Lyman-$\alpha$ radiation and collisions (i.e., $x_\alpha = x_c = 0$), the spin temperature equates with the background temperature, $T_s = T_R$, resulting in no 21-cm signal, $\Delta T_{21} = 0$. Thus, the observability of the 21-cm line critically depends on the thermal history of $T_s$. The relative coupling of the spin temperature to the kinetic gas temperature and the CMB dictates whether the 21-cm transition appears in emission, absorption, or remains unobservable.\\
The color temperature is often approximated as $T_\alpha \approx T_k$ in the cosmic dawn regime because Lyman-$\alpha$ photon scattering is dominant and coupled with the gas temperature during cosmic dawn. In that case, \eqref{eq:Ts} gives
\begin{equation}
    1 - \frac{T_R}{T_s} = \frac{x_c + x_\alpha }{1 + x_c + x_\alpha} \left(1 - \frac{T_R}{T_k} \right).
\label{eq:TsTk}
\end{equation}
Note that when $x_{\rm tot} = x_c + x_\alpha \gg 1$, the spin temperature becomes strongly coupled to the gas temperature so that $T_s \sim T_k < T_R$ and there is an absorption signal \cite{Pritchard_2008}. The collisional coupling coefficient is given by \cite{Zygelman2005, Mittal_2022}:
\begin{equation}
x_{\mathrm{c}} = \frac{T_{\ast} n_{\mathrm{H}}}{T_{R} A_{10}}
\left[ (1 - x_{\mathrm{e}}) \kappa_{\mathrm{HH}} + x_{\mathrm{e}} \kappa_{\mathrm{eH}} + x_{\mathrm{p}} \kappa_{\mathrm{pH}} \right],
\label{eq:xcoll}
\end{equation}
where $A_{10} = 2.85\times 10^{-15} \, \mathrm{s}^{-1}$ is the Einstein coefficient for spontaneous emission in the hyperfine transition, $T_{\ast} = 0.068\,\mathrm{K}$ is the equivalent temperature corresponding to the 21-cm transition energy, $n_{\mathrm{H}}$ is the number density of hydrogen atoms, and $x_{\mathrm{e}}$ is the free electron fraction whereas $x_e \simeq x_p$ for nuetral hydrogen. The rate coefficients $\kappa_{\mathrm{HH}}$, $\kappa_{\mathrm{eH}}$, and $\kappa_{\mathrm{eH}}$ depend on $T_{\mathrm{k}}$; fitting functions are provided in \cite{Mittal_2022}. The Lyman-$\alpha$ coupling coefficient is defined as:
\begin{equation}
x_{\alpha} = S_\alpha \frac{J_{\alpha}}{J_{0}},
\end{equation}
where $S_\alpha$ is called the scattering correction and the quantity $J_{0} = 5.54\times 10^{-8} (1+z) \, \mathrm{m^{-2}\,s^{-1}\,Hz^{-1}\,sr^{-1}}$ is a normalization factor determined by fundamental constants and CMB radiation~\cite{Mittal:2020kjs}. The flux $J_{\alpha}$ is the angle-averaged Lyman-$\alpha$ specific intensity, expressed as 
\begin{equation}
J_{\alpha}(z) = \frac{c}{4\pi} (1+z)^{2} 
\sum_{j=2}^{23} f(n)
\int_{z}^{z_{\mathrm{max},n}} 
\frac{ \epsilon_\alpha(\nu',z') }{ H(z') } \, dz' ,
\end{equation}
where $c$ is the speed of light, $f(n)$ are the recycling factors for Lyman-series photons, $z_{\mathrm{max},n}$ is the upper integration limit, and $\epsilon_\alpha(\nu,z)$ is the comoving emissivity, which can be computed following \cite{Pritchard_2007}. We neglect the higher cascades of the Lyman series \cite{Pritchard_2006}.\\
The evolution of the gas kinetic temperature is given by~\cite{Furlanetto2006}
\begin{equation}
\frac{dT_k}{dz} = \frac{2T_k}{1+z} - \frac{2}{3H(z)(1+z)} \sum_i \frac{\epsilon_i}{k_B n},
\end{equation}
where the first term is due to adiabatic expansion, $k_B$ is the Boltzmann constant, and $n$ is the baryon number density. The $\epsilon_i$'s represent heating contributions from: i) Heating due to Compton scattering: energy exchange between residual electrons and CMB photons~\cite{Bharadwaj_2004}
  \begin{equation}
  \epsilon_{\text{Comp}} = \frac{3}{2} n k_B \frac{x_e}{1 + x_e + f_{\mathrm{He}}} \frac{8 \sigma_T u_\gamma}{3 m_e c} (T_R - T_k),
  \end{equation}
where $f_{\rm He}$ is the fraction of Helium abundance, $\sigma_T$ represents the Thomson scattering cross-section, and $u_\gamma$ denotes the energy density of the background radiation. \\
ii) Heating due to X-ray emission: energy injection from X-ray photons emitted by galaxies and galaxy clusters can contribute to heating the IGM~\cite{mesinger2013}
  \begin{equation}
  \epsilon_X = 3.4 \times 10^{33} f_{\mathrm{heat}} f_X \left( \frac{\dot{\rho}_\star(z)}{M_\odot\, \text{yr}^{-1} \text{ Mpc}^{-3}} \right) \text{ J s}^{-1} \text{ Mpc}^{-3},
  \end{equation}
where $M_\odot$ refers to the solar mass and the parameters $f_\text{heat}$ and $f_X$ correspond to the fraction of X-ray energy deposited as heat and a normalization factor that accounts for differences between local and high-redshift observations, respectively. The Lyman-$\alpha$ flux $J_\alpha$, and consequently $x_\alpha$, depends on the star formation rate density $\dot{\rho}_\star(z)$, given as~\cite{loeb2013}
\begin{equation}
\dot{\rho}_\star(z) = -f_\star \Omega_{b,0} \rho_{c,0} (1+z) H(z) \frac{d}{dz} f_{\mathrm{coll}}(z),
\end{equation}
where $f_\star$ is the stellar-to-baryon ratio, $\Omega_{b,0}$ and $\rho_{c,0}$ are the baryon density parameter and critical density of the Universe at present, and $f_{\mathrm{coll}}(z)$ is the collapsed fraction of matter into halos, calculated via the Press-Schechter formalism~\cite{press1974}
\begin{equation}
f_{\mathrm{coll}}(z) = \mathrm{erfc} \left( \frac{1.686 (1+z)}{\sqrt{2\sigma^2(M_{\mathrm{min}})}} \right),
\end{equation}
where $\sigma^2$ is the mean squared mass variance computed from the matter power spectrum. The minimum virial temperature enters the model through the expression of minimum halo mass for star formation, i.e.,
\begin{equation}
    M_{\mathrm{min}} = \frac{10^8 h^{-1} M_\odot}{\sqrt{\Omega_{m,0}}} \left( \frac{10}{1+z} \frac{0.6}{\mu} \frac{T_{\mathrm{vir}}^\mathrm{min}}{1.98 \times 10^4 \rm{K}} \right)^{3/2}.
\end{equation}
Here we adopt the following fiducial parameters:
$f_{\mathrm{X}}=1$, $f_{\mathrm{heat}}=0.2$, $f_{\star}=0.03$, $\mu = 1.22$, and
$T_{\mathrm{vir}}^\mathrm{min}=10^3\,\mathrm{K}$ \cite{BarkanaLoeb2001}.\\
iii) Heating due to PMF dissipation: PMFs can directly heat the IGM by transferring energy from the magnetic sector to the kinetic sector of the baryons. The two dominant dissipation mechanisms after recombination are ambipolar diffusion (AD) and decaying turbulence (DT). The heating rate due to ambipolar diffusion is given by~\cite{Sethi:2004pe, Chluba:2015lpa, Bera:2020jsg}
\begin{align}
\epsilon_{\rm AD} &=
\frac{1 - x_e}{\gamma x_e\; \rho_b^2(z)}\;
\frac{\big\langle \,|(\nabla \times \mathbf{B}) \times \mathbf{B}|^2 \,\big\rangle}{16\pi^2},
\label{eq:AD_general}
\end{align}
where $\rho_b$ is the baryon mass density at redshift $z$, the ion–neutral coupling coefficient $\gamma = 1.94\times 10^{11} (T_k/{\rm K})^{0.375}
\;\;{\rm m^3\,kg^{-1}\,s^{-1}}$~\cite{Schleicher:2008aa}, and the average squared Lorentz force term can be estimated as~\cite{Chluba:2015lpa}
\begin{align}
\frac{\big\langle \,|(\nabla \times \mathbf{B}) \times \mathbf{B}|^2 \,\big\rangle}{16\pi^2 }
\simeq \frac{\rho_B^2(z)}{\,\ell_d^2(z)}\,f_L(n_B+3).
\label{eq:Lorentz_sq}
\end{align}
The magnetic field energy density is $\rho_B(z) = \frac{|\mathbf{B}|^2}{2 \mu_0}$, the physical damping scale is related to the comoving damping wavenumber $k_D$ by $\ell_d(z) = \frac{1}{(1+z)k_D}$ where $k_D$ is given by (\ref{kD}), and the dimensionless factor $f_L$, which depends on the PMF spectral index $n_B$, is given by
\begin{align}
f_L(n_B+3) = 0.8313\left[1 - 1.02\times 10^{-2}\,(n_B+3)\right](n_B+3)^{1.105}.
\end{align}
The heating rate due to decaying magnetohydrodynamic turbulence is~\cite{Sethi:2004pe}
\begin{align}
\epsilon_{\rm DT} &=
\frac{3m}{2}\;
\frac{\Big[\ln\!\big(1+t_d/t_i\big)\Big]^m\;}
{\Bigg[\ln\!\big(1+t_d/t_i\big) + \frac{3}{2}\ln\!\left(\frac{1+z_i}{1+z}\right)\Bigg]^{m+1}}
\,H(z)\,\rho_B(z),
\label{eq:DT_general}
\end{align}
where $m = \frac{2(n_B+3)}{n_B+5}$, the initial epoch $t_i$ (the corresponding redshift $z_i$) when the decay starts and the physical decay timescale $t_d$ for the turbulence are related by $t_d/t_i \simeq 14.8\left(\frac{B_0}{\rm nG}\right)^{-1}\left(\frac{k_D}{\rm Mpc^{-1}}\right)^{-1}$~\cite{Chluba:2015lpa}.\\
It is important to note that ambipolar diffusion and decaying turbulence heating vanish in the case of a scale-invariant primordial magnetic field (PMF) spectrum, $n_B = -3$. Physically, this means that a scale-invariant PMF does not inject energy into the baryonic gas after recombination. Dissipative heating requires excess magnetic power on small scales, but in the scale-invariant case the spectrum is flat in $k$-space and provides no preferential small-scale structure to drive ambipolar diffusion or turbulent decay. As a result, the IGM does not experience significant PMF heating in this special case.

%%%%%%%%%%%%%%%%%%%%%%%%%%%
\section*{Acknowledgement}
%%%%%%%%%%%%%%%%%%%%%%%%%%%
We thank Debarun Paul, Arko Bhaumik and Sourav Pal for fruitful discussions. We acknowledge the computational facilities of the Technology Innovation Hub, Indian Statistical Institute (ISI), Kolkata. We also acknowledge the use of \texttt{Mathematica} for the analytic and numerical computations, and \texttt{CLASS} for the cosmological analysis carried out in this work. S thanks ISI, Kolkata, for financial support through a Junior Research Fellowship. MRH thanks ISI, Kolkata, for financial support through a Research Associateship. RK thanks SP for his kind support during the visit to ISI, Kolkata, where this work was initially planned, and gratefully acknowledges the financial support from Shanghai University through the Postdoctoral Fellowship. SP thanks the ANRF, Govt. of India, for partial support through Project No. CRG/2023/003984.

%%%%%%%%%%%%%%%%%%%%%%%%%%%%%%%%%%%%%%%%%%%%%%%%%
\bibliographystyle{JHEP}
\bibliography{Ref}
\end{document}